\documentclass[twocolumn]{aastex62}

\usepackage{graphicx}
\usepackage{txfonts}
\usepackage{hyperref}

\shorttitle{Origin of a bipolar magnetic region}
\shortauthors{Getling and Buchnev}

\begin{document}

\title{The origin and early evolution of a bipolar magnetic region in the solar photosphere}

\correspondingauthor{A. V. Getling}
\email{A.Getling@mail.ru}

\author{A. V. Getling}
\affil{Skobeltsyn Institute of Nuclear Physics, Lomonosov Moscow State
              University, Moscow, 119991 Russia}

\author{A. A. Buchnev}
\affil{Institute of Computational Mathematics and Mathematical Geophysics,
             Novosibirsk, 630090 Russia}

 \begin{abstract}
     Finding the formation mechanisms for bipolar configurations of strong local magnetic field under control of the relatively weak global magnetic field of the Sun is a key problem of the physics of solar activity.  This study is aimed at discriminating whether the magnetic field or fluid motion plays a primary, active role in this process.  The very origin and early development stage of Active Region 12548 are investigated based on SDO/HMI observations of 2016 May 20--25. Full-vector magnetic and velocity fields are analyzed in parallel. The leading and trailing magnetic polarities are found to grow asymmetrically in terms of their amplitude, magnetic flux, and the time variation of these quantities. The leading-polarity magnetic element originates as a compact feature against the background of a distributed trailing-polarity field, with an already existing trailing-polarity magnetic element. No signs of strong horizontal magnetic fields are detected between the two magnetic poles. No predominant upflow between their future locations precedes the origin of this bipolar magnetic region (BMR); instead, upflows and downflows are mixed, with some prevalence of downflows. Any signs of a large-scale horizontal divergent flow from the area where the BMR develops are missing; in contrast, a normal supergranulation and mesogranulation pattern is preserved. This scenario of early BMR evolution is in strong contradiction with the expectations based on the model of a rising $\Omega$-shaped loop of a flux tube of strong magnetic field, and an \emph{in situ} mechanism of magnetic-field amplification and structuring should operate in this case.
\end{abstract}

\keywords{solar active region magnetic fields --- solar active region velocity fields --- solar photosphere --- sunspots
               }
\section{Introduction}

The origin of active regions (ARs) and bipolar sunspot groups is among the key issues to be resolved to comprehend the nature of solar activity. In essence, the central problem is finding a mechanism for the formation of bipolar configurations of strong magnetic field under control of the relatively weak global magnetic field of the Sun; such bipolar magnetic regions (BMRs), giving rise to sunspot groups, trigger the whole sequence of active processes over a wide range of heliocentric distances.

In the process of sunspot formation, a primary role may be played by either magnetic field or plasma motion. In the first case, the magnetic field, having achieved a high strength before the initiation of the sunspot-forming process, proves to be able to dictate one type of solar-plasma motion or another exerting magnetic forces on the matter. In the second case, plasma motion itself produces a strong magnetic field and imparts a bipolar configuration to it according to the laws of magnetohydrodynamics.

The first situation is assumed, in particular, by the widely known \emph{rising-tube model} (RTM), which attributes the formation of a bipolar sunspot group to the emergence of an $\Omega$-shaped loop of a coherent flux tube of strong magnetic field (by the RTM, we mean the physical view of the process rather than the computational thin-flux-tube model). The second situation is characteristic of various possible mechanisms of \emph{in situ} magnetic-field \emph{amplification and structuring} due to plasma flows, e.g., convection; some of these mechanisms can be classified as local dynamos. To approach the understanding of the sunspot-formation processes, it is of great importance to discriminate between these two possibilities.

According to the RTM, the general toroidal\footnote{As is typical of the
literature on stellar and planetary dynamos, we use here the terms
\emph{toroidal} and \emph{azimuthal} as synonyms, although they are not
mathematically equivalent.} magnetic field produces a strong flux tube deep in the convection zone, down to the tachocline, whereupon a loop of the tube is formed and then lifted by the magnetic-buoyancy force \citep[whose role was first recognized by][]{parker}. The RTM agrees well with such important regularities of solar activity as Hale's polarity law and Sp\"orer's law of sunspot-formation latitudes. For this reason, the properties of the rising tube have become the object of numerous studies \citep[see, e.g.,][etc.] {clgretal1,clgretal2,fanetal,rempcheung}. \citet{Fan2009} reviewed studies of the conceivable  processes of magnetic-flux-tube rising, giving primary attention to both thin-flux-tube model calculations (which fail at depths of 20--30~Mm, where the cross-sectional size of the tube becomes comparable with the local scale height) and full 2D or 3D numerical MHD simulations based on nonlinear equations for a compressible fluid. Most of these studies consider initially present tubes without discussing the process of their formation. In particular, \citet{JouveBrun2009}, dealing with a spherical geometry and using the anelastic approximation, simulated latitudinally stretched, initially axisymmetric magnetic flux tubes rising in a rotating turbulent convection zone from its base and fragmenting; interaction of the tubes with convection and large-scale flows was also considered.

Studies aimed at describing the formation of flux tubes as a result of the instability of a magnetic layer \citep{Fan2001} and the formation of the magnetic layer itself in a velocity-shear layer \citep{Vasil_Brummell_2008} are not numerous. They gave no definite indications for these possibilities under the conditions of the solar convection zone.

Since a twist stabilizes the tube, maintaining its cohesion, and in view of the observed twist of the AR magnetic fields, the rising tube is typically assumed to be twisted. Some analyses of the magnetic fields observed in ARs, with determinations of the magnetic helicity, were carried out with this idea behind \citep{Luoni_etal:2011,Poisson_etal:2015}.

The RTM was considered a standard paradigm in the studies of AR-formation processes for several decades. In recent years, however, abundant observational data of very high spatiotemporal resolution have progressively cast more and more doubts upon the universal adequacy of this model.

As can easily be imagined, the emergence of an $\Omega$-shaped loop of strong-magnetic-flux tube should entail three striking observable effects, viz.:
\begin{enumerate}
   \item An upflow between the two future magnetic poles of the BMR, on a scale of no less than the distance between them.
   \item Strong horizontal magnetic fields at the apex of the emerging flux-tube loop.
   \item Intense spreading of matter from the loop-emergence site on the scale of the entire BMR.
\newcounter{lenumi}\setcounter{lenumi}{\value{enumi}}
\end{enumerate}
As we will see, there is no convincing observational evidence for the actual presence of these effects. Nevertheless, some facts can be interpreted in terms of features 1 and~3.

\citet{Grigor'ev_etal2007} report an enhanced plasma upflow preceding the formation of a new magnetic configuration in the developing AR~10488. In their opinion, this upflow can be attributed to the flux-tube-rising process. Let us note, however, that the Doppler-velocity and magnetic-field patterns presented by these authors do not seem to be spatially correlated in a way typical of such a process.

In their MHD simulations of flux-tube emergence, \citet{Toriumi_Yokoyama2012, Toriumi_Yokoyama2013} arrived at the quite expectable conclusion that a horizontal divergent flow (HDF) should precede the appearance of the magnetic flux. \citet{Toriumi_etal_2012} observationally detected signatures of HDFs prior to the magnetic-flux emergence. \citet{Khlystova_Toriumi_2017}, using SOHO/MDI observations of the emergence of small AR 9021 and AR 10768, found strong upflows on a mesogranular scale at the initial stage of active-region formation. They noted good agreement in the time variation of the plasma-upflow velocity and area between these observations and numerical simulations of flux-tube emergence carried out by \citet{Toriumi_etal_2011}. Strong HDFs in a number of emerging ARs were also revealed by \citet{Khlystova2013a} and \citet{Toriumi_etal_2014} on the basis of SOHO/MDI observations. In these studies, Doppler measurements were carried out away from the disk center to determine the horizontal velocities by properly projecting the line-of-sight velocities. Although the horizontal velocity can be determined in this way more accurately at larger distances from the disk center, it should be kept in mind that the resolution of the velocity pattern on the solar surface degrades with this distance. Moreover, the discrimination between the spread velocity related to the AR development and the regular supergranulation flow is a particular, not simple task.{\sloppy

}There are, however, observational facts definitely contradicting the above-mentioned features of the tube-rising process. In particular, \citet{PevtsovLamb:2006} ``observed no consistent plasma
flows at the future location of an active region before its
emergence,'' and \citet{kosov} detected no ``large-scale flow patterns on the surface, which would indicate emergence of a large flux-rope structure''; instead, local updrafts and downdrafts were observed. As shown by \citet{Birch_etal_2016}, the velocity fields around emerging BMRs are statistically very similar to the velocity fields in the quiet-Sun photosphere in terms of the presence of HDFs (we will return to this finding in Subsection~\ref{velfield}).

A further example of the AR-development pattern at an early formation stage of a new BMR within already existing AR 11313 was given by \citet{getling_etal_hinode, getling_etal_hinode_2} (hereinafter, Papers~I and II, respectively). Neither a horizontal spreading on the scale of the whole developing subregion, nor a strong horizontal magnetic field between the growing sunspots, nor a strong upflow at that site was detected. Thus, a noticeable discrepancy was found between the observed evolutionary scenario of the magnetic and velocity fields and the RTM-based expectations.

Doubts about the adequacy of the rising-tube model are based not only on the absence of convincing observational evidence for effects 1--3 but also on the following:
\begin{enumerate}\addtocounter{enumi}{\value{lenumi}}
\item No quite satisfactory explanation of the origin of a coherent flux tube of strong magnetic field deep in the convection zone has been suggested. The known hypotheses differ in their plausibility and the appropriateness of their starting points \citep[see, e.g., the already mentioned works:][]{Fan2001,Vasil_Brummell_2008}. It is important that an intense flux tube should affect the structure of the convective velocity field before the emergence on the photospheric surface; such an influence is not actually observed.
\item The tilt of sunspot groups is typically interpreted as an effect of the Coriolis force on the emerging $\Omega$-shaped flux-tube loop. Therefore, the tilt should be the smaller, the stronger the magnetic field counteracting the turning of the loop. However, as \citet{kosovstenflo} and \citet{kosov} note, observations do not demonstrate such a magnetic-flux dependence of the tilt angle; their ``study of the variations of the tilt angle of bipolar magnetic regions (BMRs) during the flux emergence questions the current paradigm that the magnetic flux emerging on the solar surface represents large-scale magnetic flux ropes ($\Omega$-loops) rising from the bottom of the convection zone.''
\end{enumerate}

   \begin{figure*}
   \centering
   \includegraphics[width=0.35\textwidth,bb=0 0 504 360] {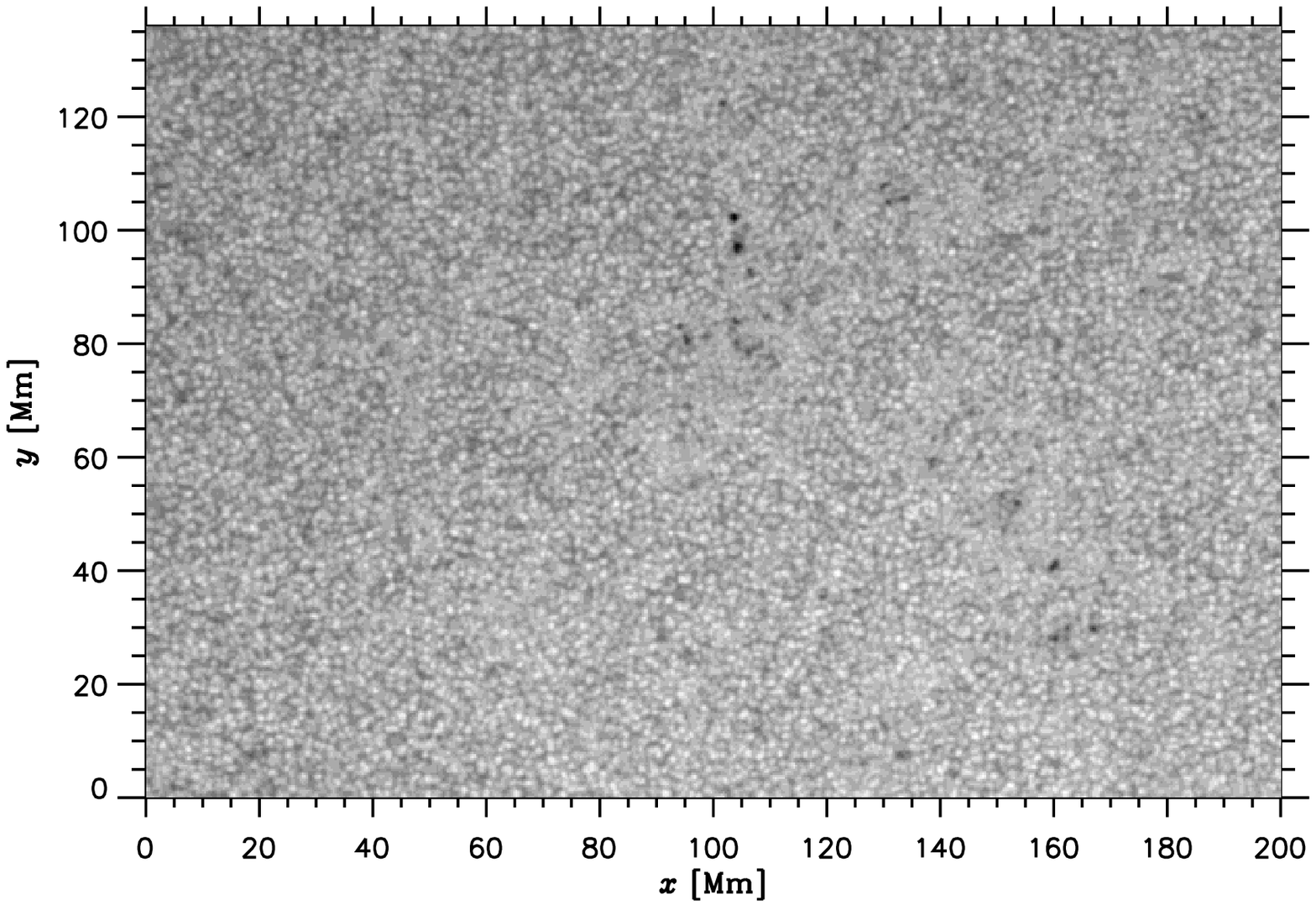}
   \includegraphics[width=0.35\textwidth,bb=0 0 504 360] {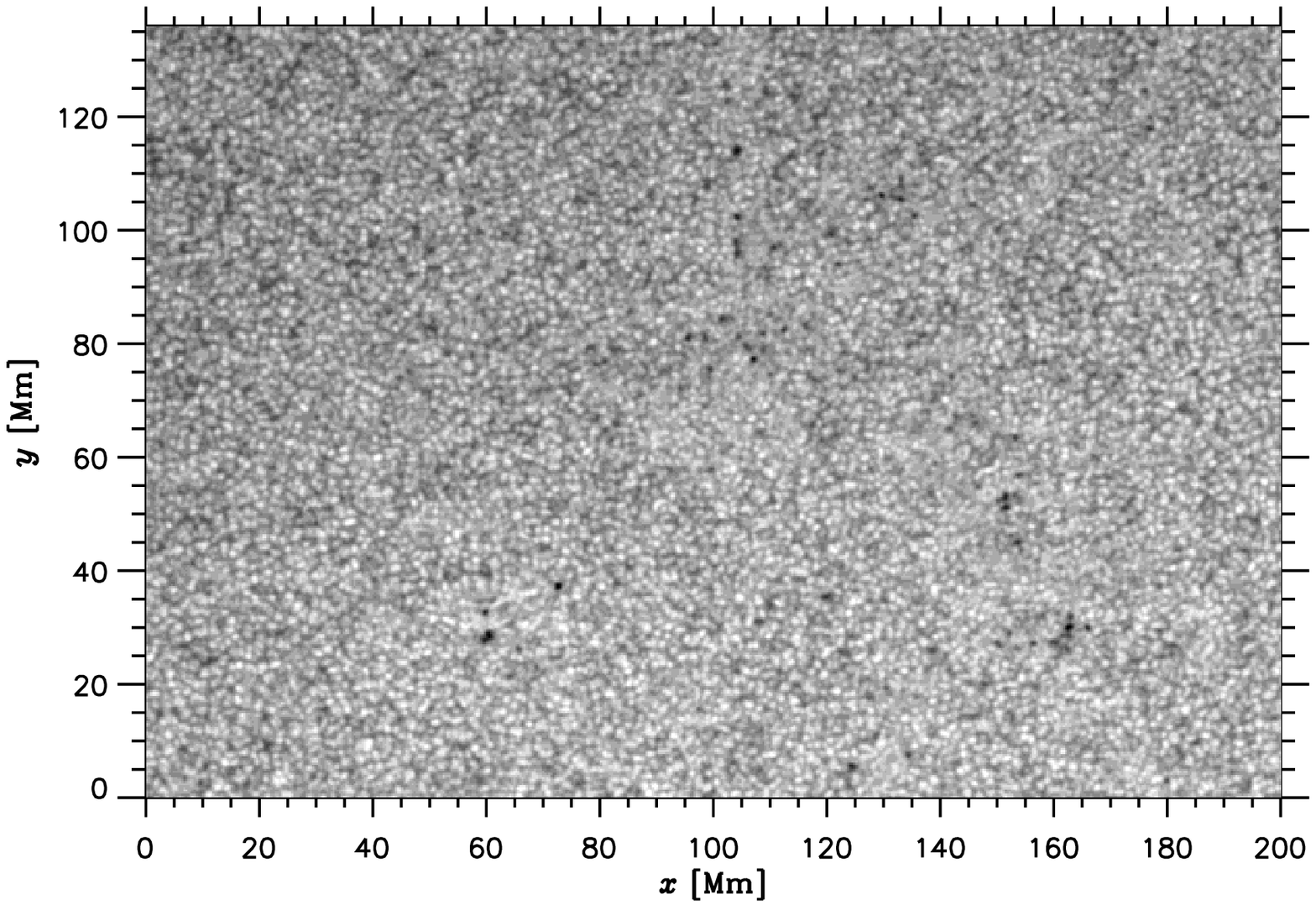}\\
   \tiny{\hspace{0.5cm}2016 May 23, 20:00 TAI (RT)\hspace{4.cm} 2016 May 23, 23:00 TAI}\\    
   \includegraphics[width=0.35\textwidth,bb=0 0 504 360] {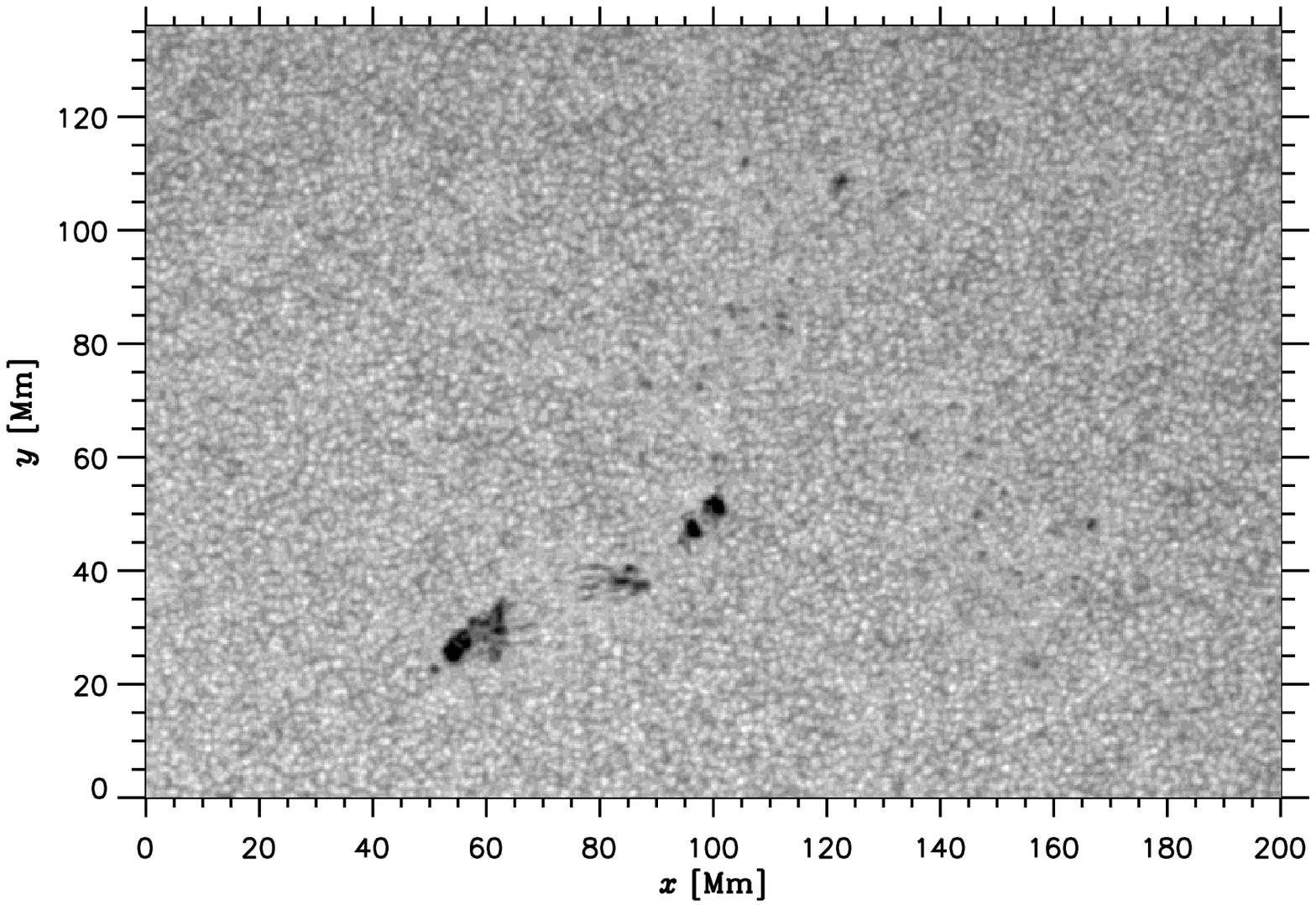}
   \includegraphics[width=0.35\textwidth,bb=0 0 504 360] {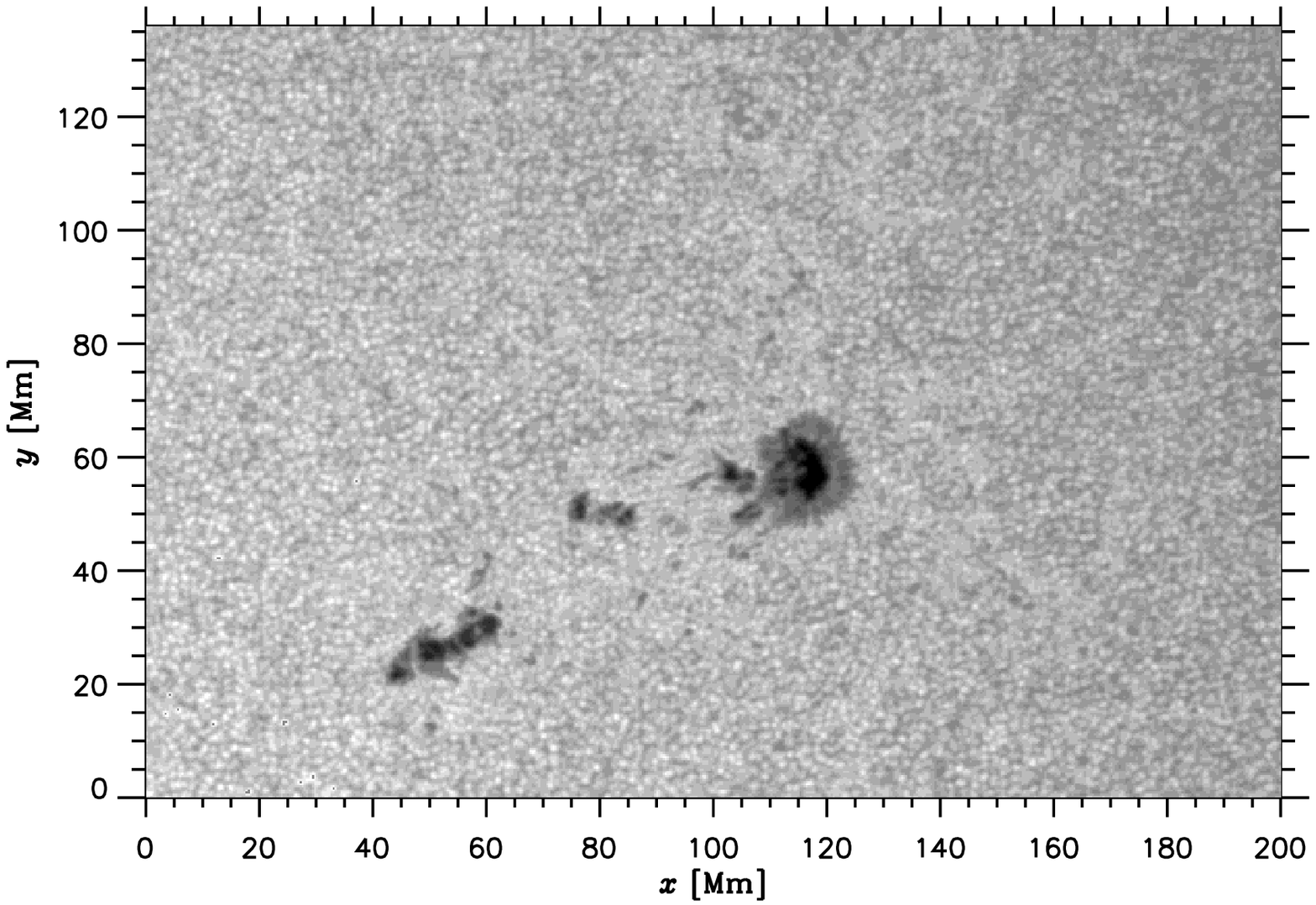}\\
   \tiny{\hspace{0.7cm}2016 May 24, 20:00 TAI \hspace{4.2cm} 2016 May 25, 20:00 TAI}
   \caption{Evolution of AR 12548. White-light SHARP images are shown
                for the times indicated under each of them.\protect}
              \label{white}%
    \end{figure*}

Among others, \citet{warnetal2013, Jabbari_etal2016, warneckeetal2016} critically discuss the appropriateness of the RTM. Their reasoning is based not only on the data of immediate observations but also on helioseismological inversions and direct numerical simulations. In particular, they remark that no signs of rising flux tubes have yet been found in helioseismology. These researchers treat the formation of sunspots as a shallow phenomenon and investigate the possible role of the so-called negative-effective-magnetic-pressure instability (NEMPI; the negative pressure is due to the suppression of the total turbulent pressure -- the sum of hydrodynamic and magnetic contributions -- by the magnetic field).

As alternatives to the RTM, various mechanisms of local \emph{(in situ)} magnetic-field amplification and structuring have been suggested. Among them are a hydromagnetic instability related to quenching of eddy diffusivity by the enhanced magnetic field and cooling-down of the plasma \citep{kitmaz}, the already mentioned NEMPI, and various MHD mechanisms
of inductive excitation of magnetic fields strongly coupled with fluid motions
\citep[\emph{local dynamos;} see, e.g., numerical simulations by][in which the initial presence of a uniform, untwisted, horizontal magnetic field is assumed]{stein_nord2012}. In particular, based on both observations and theory, \citet{Cheung_etal2017} note that the convective dynamo should operate in the convection zone over various spatial scales, without a clear separation between the large and small scales. We discussed some local formation mechanism for BMRs and sunspots in Paper I (and briefly in Paper II).

The vulnerability of the view of BMR origin as the emergence of a strong coherent flux tube is even reflected in the currently used terminology: the expression ``flux-tube emergence'' is now usually replaced with ``flux emergence.'' A comprehensive review of possible flux-emergence processes is given by \cite{Cheung_Isobe2014}.

Nevertheless, many researchers still consider the RTM to be a plausible mechanism of BMR formation. Extensive analyses of numerous ARs from the standpoint of discerning various possible evolutionary scenarios are important.

We study here, on a qualitative level, the very origin and early evolutionary stage of a BMR and a sunspot group in AR 12548. In contrast to the content of Papers I and II, we now consider a ``naked'' emergence of an AR \citep[i.e, after][``the flux emergence that is isolated from and unrelated to pre-existing magnetic activity'']{Centeno2012}. Our approach is based on the parallel consideration of the full-vector magnetic and velocity field in the growing BMR. The time cadence of the data used is 12~min, so that we are able to keep track of the process under a temporal ``magnifying glass.'' We will basically discuss the observed scenario in the context of its affinity with the above-mentioned implications of the tube-rising process, 1--3. As it will be seen, the development of AR 12548 appears to be strongly dissimilar to the RTM scenario. In general, the formation mechanism must not necessarily be unique for all ARs. We consider verifying the adequacy of the RTM for a wider set of ARs to be our ``tactical'' aim, which could naturally be a step toward solving the ``strategic'' problem of understanding the mechanism (or mechanisms) of sunspot formation.\\\\

\section{Observations and Data Processing}\label{obs}

We use here data from the Helioseismic and Magnetic Imager (HMI) of the Solar Dynamics Observatory (SDO), which are stored at and available from the Joint Science Operations Center (JSOC, \url{http://jsoc.stanford.edu}). A BMR that gave rise later to a sunspot group in AR 12548 originated on 2016 May 23 near the central meridian. Diffuse magnetic fields around the future BMR location were observed since their emergence at the eastern limb on May 16. The early development stage of the BMR that we will analyze here fell on May 23, and the sunspot-group formation was mainly completed by May 27.

Our analysis of the magnetic fields is based on a Spaceweather HMI Active Region Patch \citep[SHARP; see][]{Bobra_etal_2014} with the data remapped to a Lambert cylindrical equal-area projection (CEA). This automatically selected patch is centered at the flux-weighted centroid of the AR. The magnetic-field vector is decomposed into a radial (vertical), latitudinal and longitudinal components. The data used to determine the velocity-field vector were taken for an area of a size specified by us, centered at the same point. The Dopplergrams are also CEA-remapped but not projected, still representing the line-of-sight, rather than radial, velocity component (we neglect the projection effects taking advantage of the fact that the BMR was not far from the disk center on May 23, and the difference between the line-of-sight and radial, or vertical, component is not important at the moment). We compute the horizontal velocities from a series of white-light images of the same CEA-remapped area using a modified local-correlation-tracking (LCT) technique \citep{gbuch}.

The pixel size is 0.5 arcsec $\approx$ 366 km. The SHARP under study measures 547 $\times$ 372 pixels, or 200 $\times$ 136 Mm$^2$, and the size of the area used for velocity determinations is 300 $\times$ 300 pixels, or 109.8 $\times$ 109.8 Mm$^2$.

\begin{figure*}
\centering
\includegraphics[width=0.346\textwidth,bb=23 5 655 400, clip] {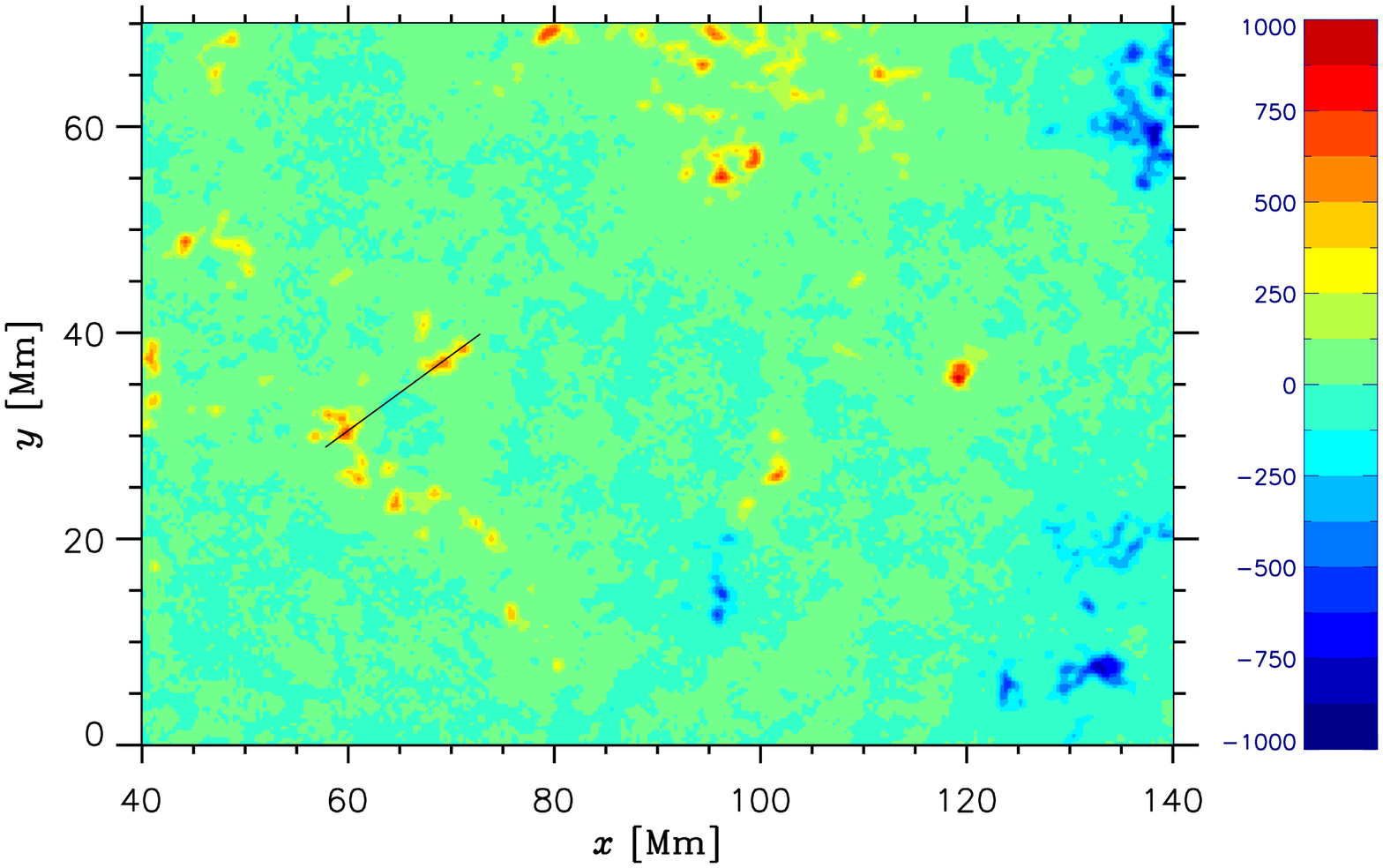}\qquad
\includegraphics[width=0.31\textwidth,bb=0 -4 485 340,clip]
{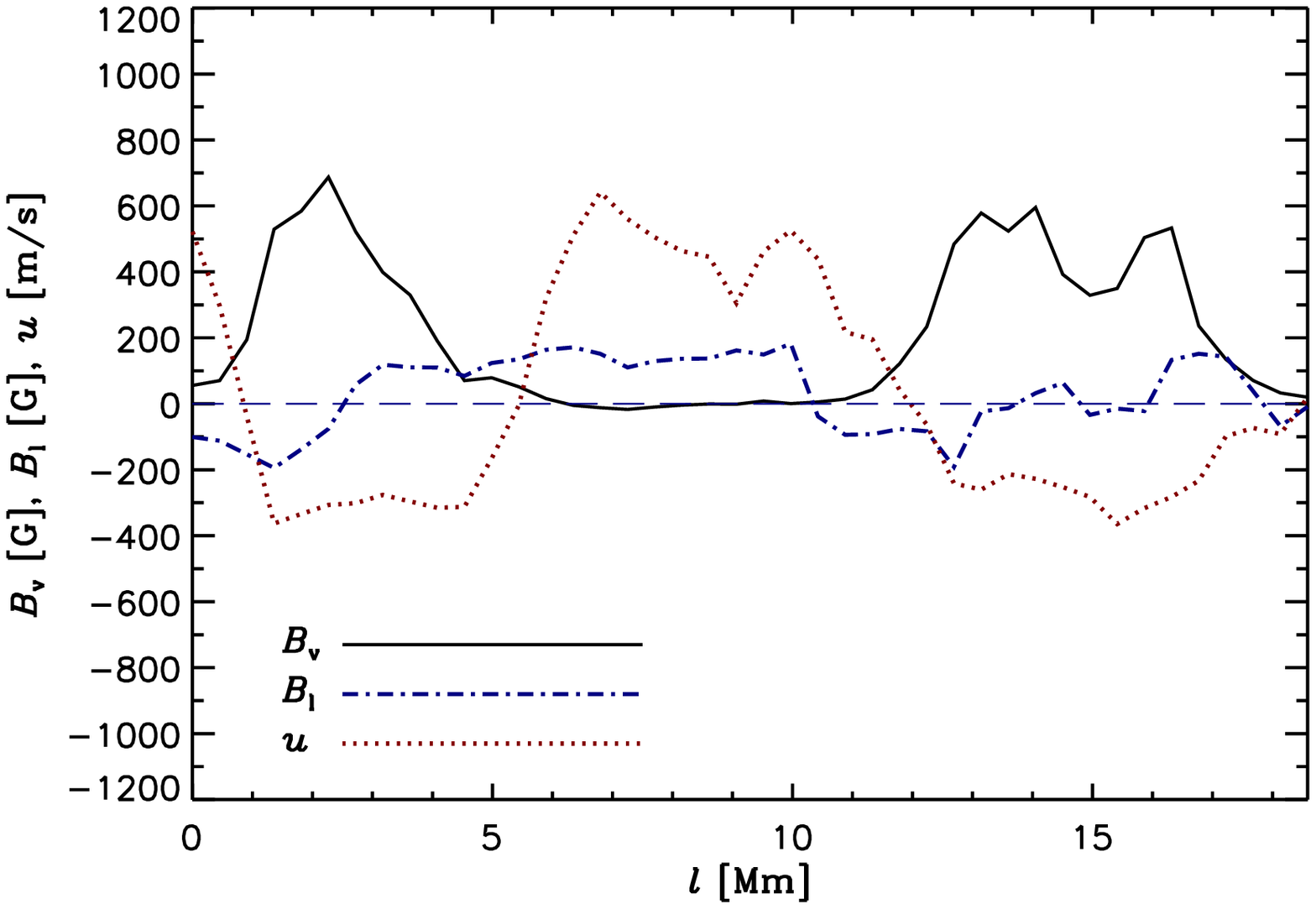}\\
\vspace{-6pt}\tiny{2016 May 23, 20:00 TAI (RT)}\\    
\includegraphics[width=0.351\textwidth,bb=35 5 550 350, clip]
{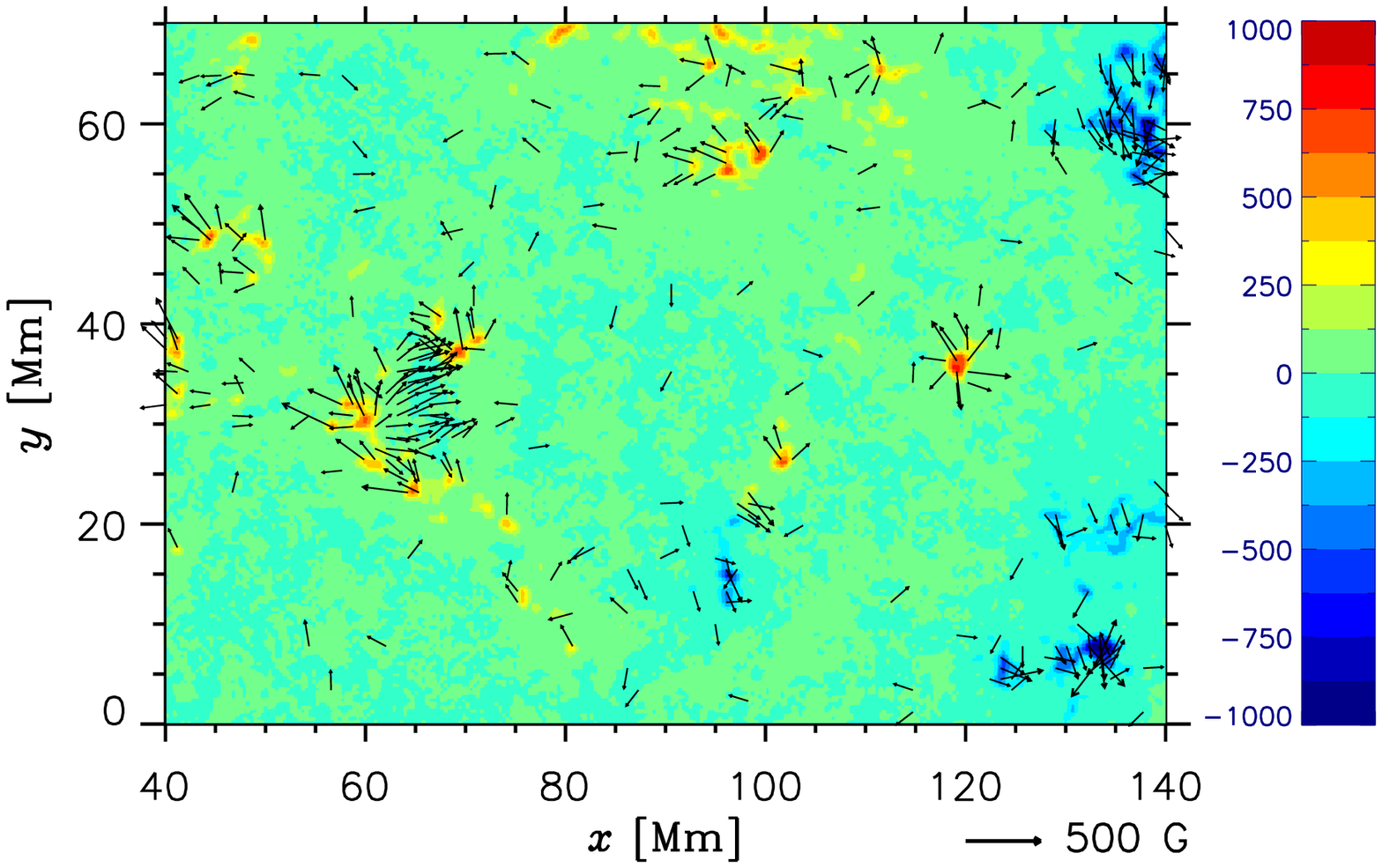}\qquad
\includegraphics[width=0.31\textwidth,bb=0 -17 485 340,clip]
{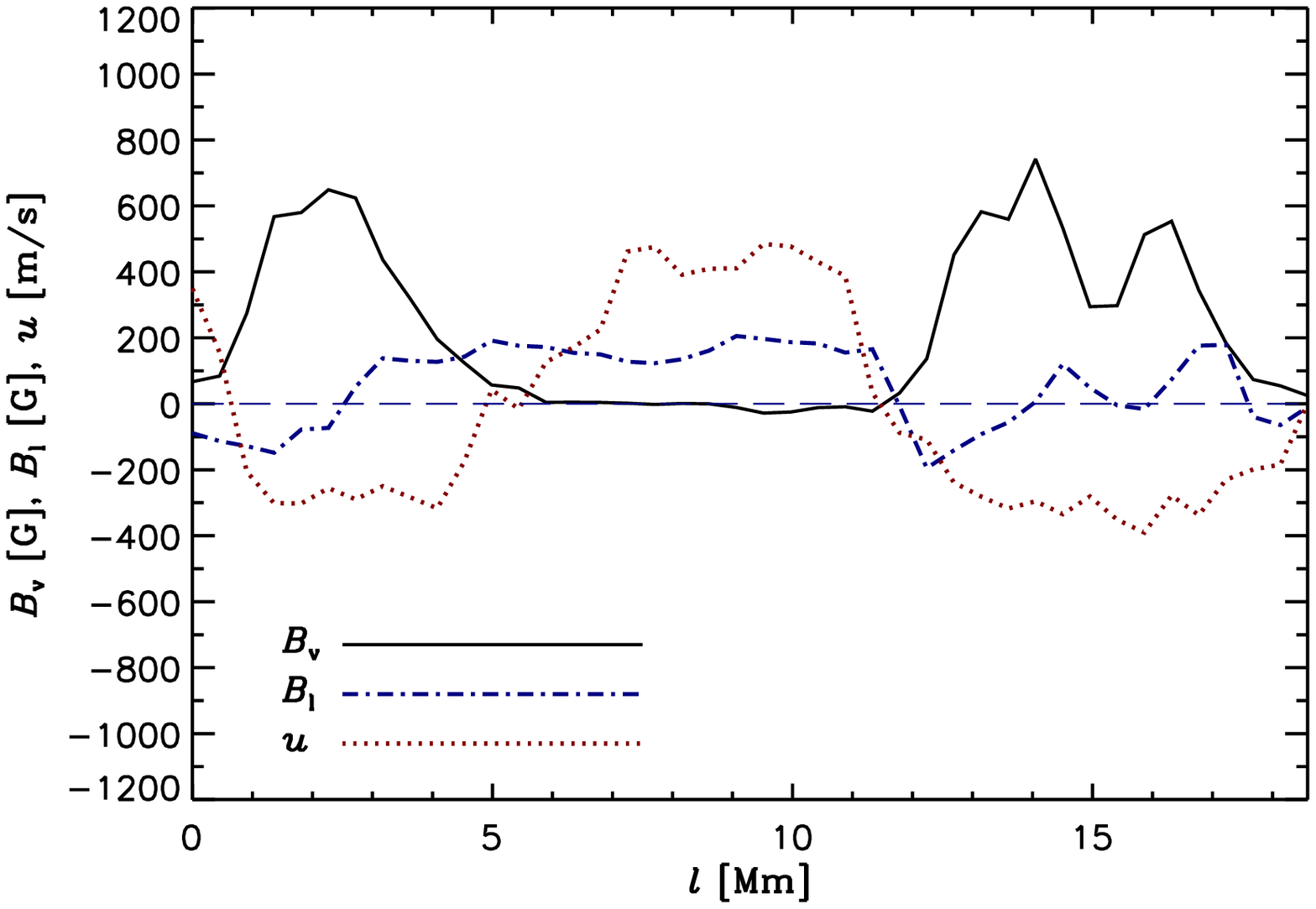}\\[4pt]
\vspace{-6pt}\tiny{2016 May 23, 20:12 TAI}\\
\includegraphics[width=0.351\textwidth,bb=35 5 550 350, clip] {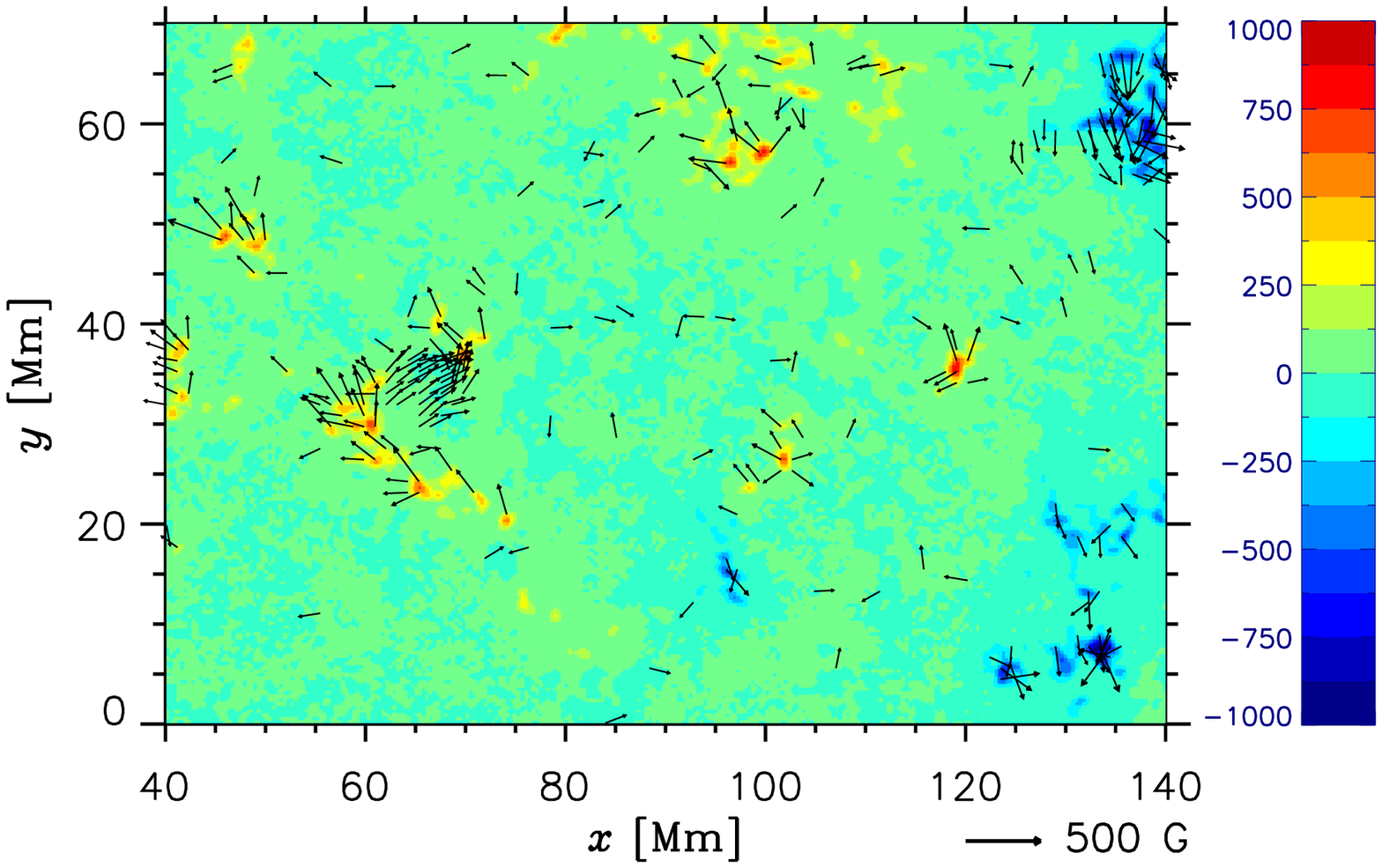}\qquad
\includegraphics[width=0.31\textwidth,bb=0 -17 485 340,clip]
{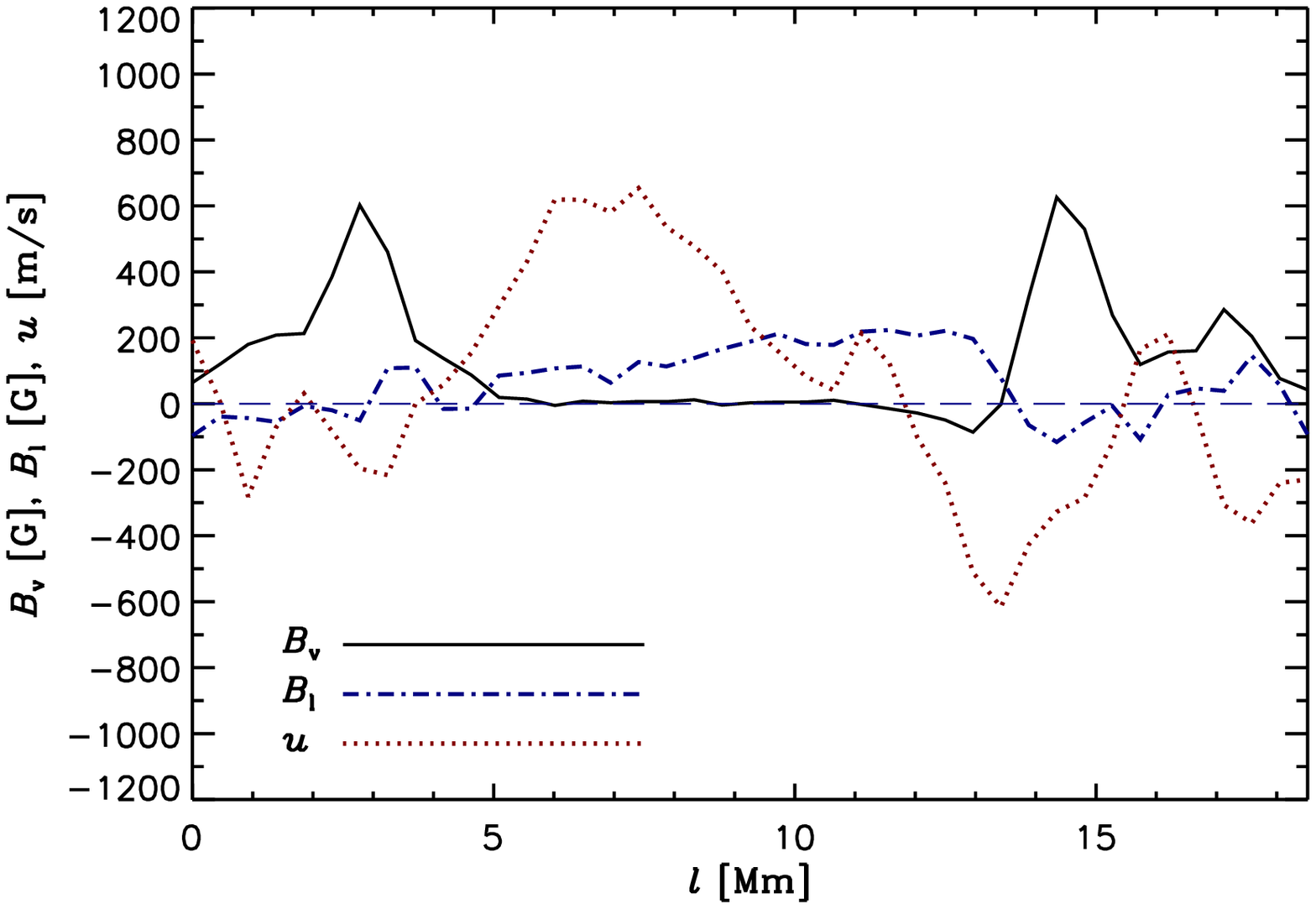}\\[4pt]
\vspace{-6pt}\tiny{2016 May 23, 20:48 TAI}\\
\includegraphics[width=0.351\textwidth,bb=35 5 550 350, clip] {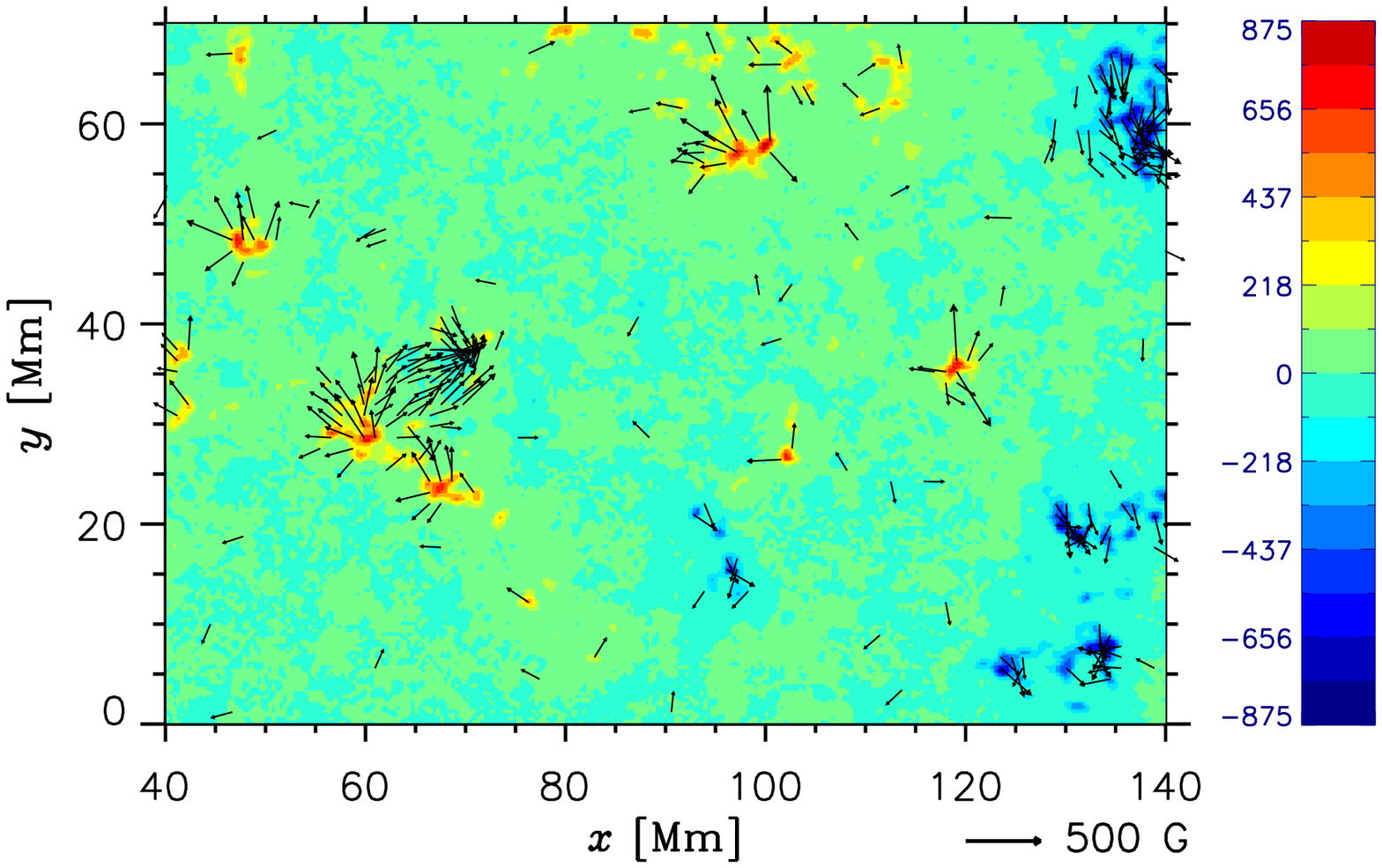}\qquad
\includegraphics[width=0.31\textwidth,bb=0 -17 485 340,clip]
{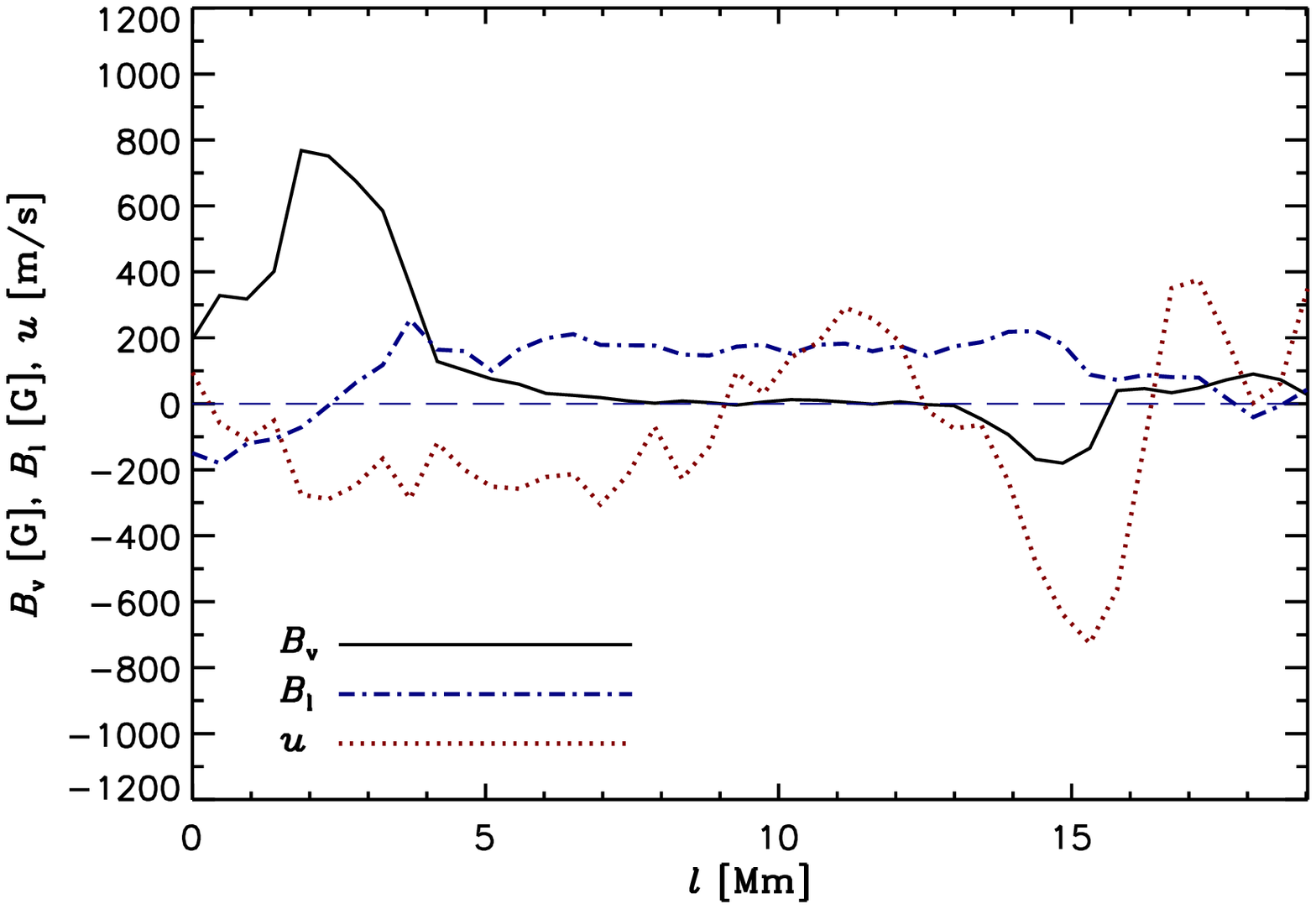}\\[4pt]
\vspace{-6pt}\tiny{2016 May 23, 21:48 TAI}\\
\includegraphics[width=0.351\textwidth,bb=35 5 550 350, clip] {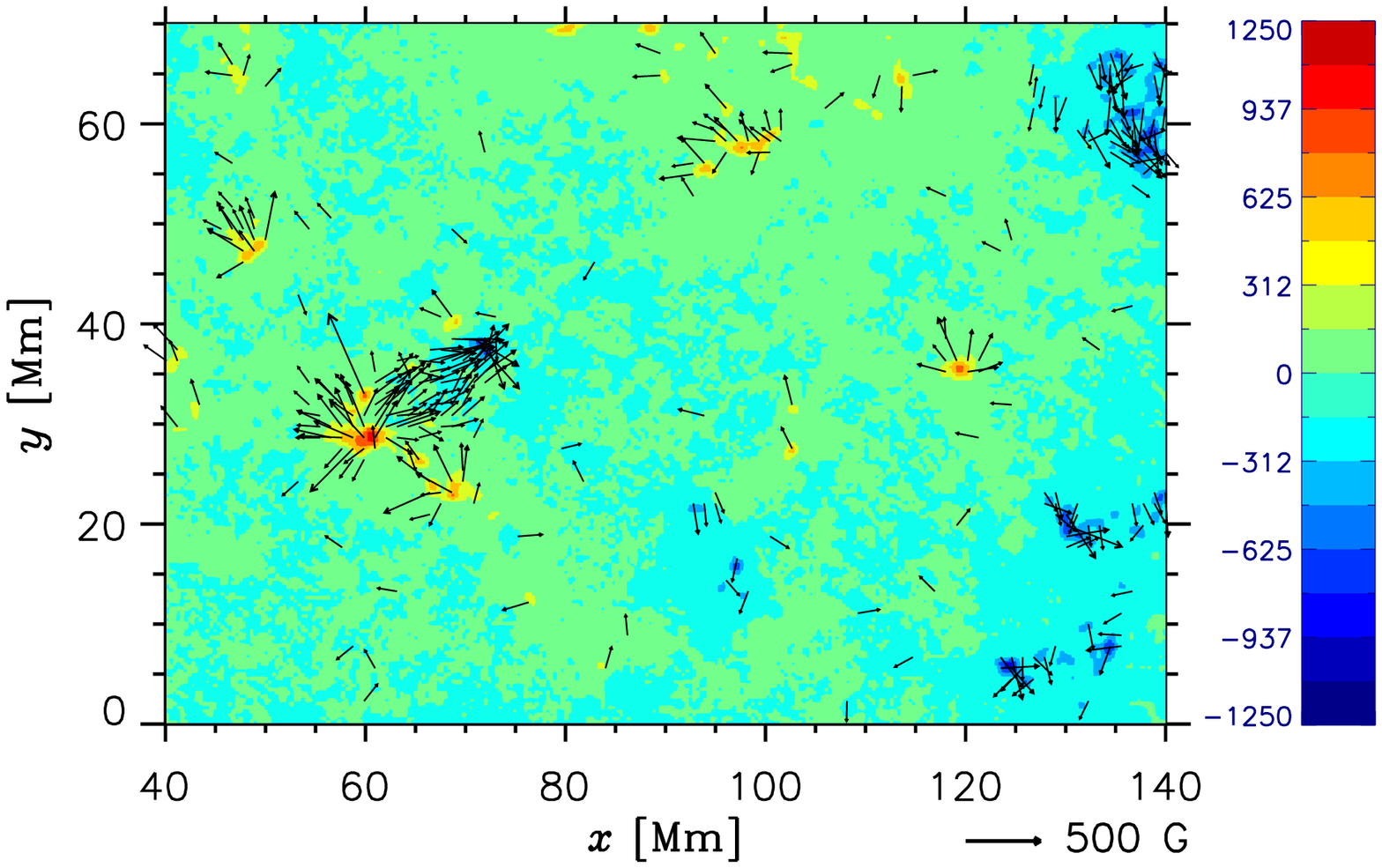}\qquad
\includegraphics[width=0.31\textwidth,bb=0 -17 485 340,clip]
{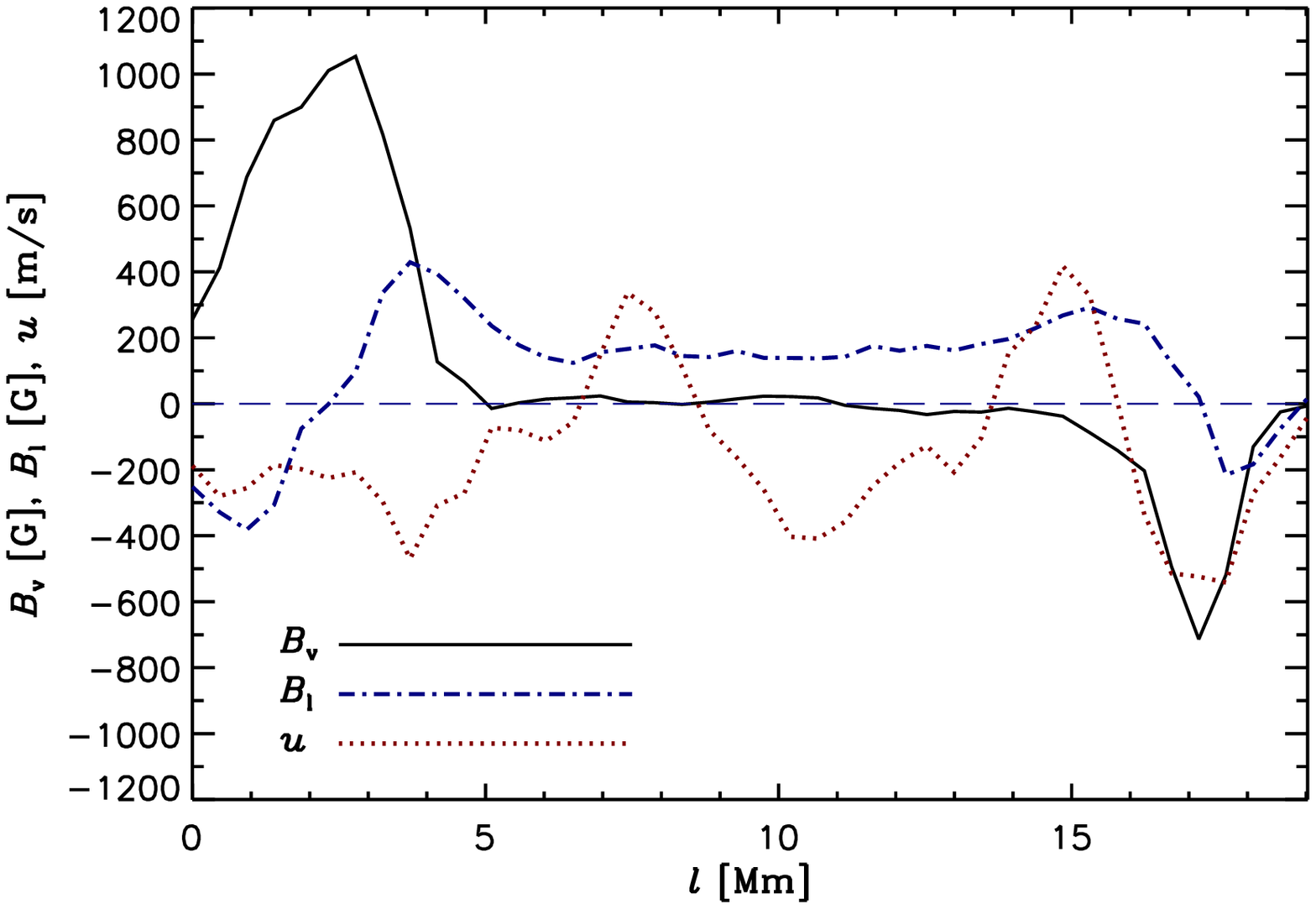}\\[4pt]
\vspace{-6pt}\tiny{2016 May 23, 22:48 TAI}\\[-4pt]
\caption{Origin of the bipolar magnetic-field structure. Left: the vertical-magnetic-field map for the RT and magnetic-field-vector maps for four subsequent times (colours representing the vertical and arrows representing the horizontal magnetic field; only vector values exceeding 150~G are shown); right: \textbf{profiles of variation along the BMR axis for the vertical magnetic field, longitudinal magnetic field and vertical velocity.} Both are given for the times indicated under each row of plots. \textbf{For each time, a line segment drawn approximately through the centroids of the main magnetic elements is assumed as the BMR axis (such a segment is shown in the first map).}}
\label{images_vs_profiles}
\end{figure*}

We applied Fourier subsonic filtering \citep{Title_etal_1989} with a cutoff phase velocity of 4 km\,s$^{-1}$ to the continuum images and Dopplergrams taken with a cadence of 45~s.  To eliminate the velocity fluctuations on a granular scale, we smoothed the line-of-sight velocities and reduced each smoothed Dopplergram to zero average.

The LCT procedure was applied to a sequence of images with a cadence of 135~s. For this procedure to be successful, we magnified the images doubling the number of pixels in each horizontal dimension with the use of a standard subroutine based on bilinear interpolation. To obtain final representations of the horizontal-velocity-vector field, we either averaged the measured velocities over nine time steps (20~m~15~s) or integrated the displacements of imaginary corks distributed over the area of interest, thus constructing cork trajectories for time intervals of 2 to 4 hours.\\

\section{Results}

\subsection{Evolution in White Light}

As a reference time (RT) for the data series that we analyze, we assume the time 2016 May 23, 20:00 TAI, when the last SHARP magnetogram without signs of the growing BMR was obtained in the 12-min-cadence series. The white-light SHARP images (Figure~\ref{white}) show that, while the photosphere in the lower left quadrant of the patch seems completely unperturbed at the RT, two clear-cut pores are present 3~h later. During the first two days starting from the RT, the sunspot group originates and acquires an appearance typical of bipolar groups, with a well-defined umbral--penumbral structure of the leading and trailing spots. At later times, the structure of the group becomes more complex and less ordered; we will not consider here these development stages.

\begin{figure*}
\centering
\includegraphics[width=0.35\textwidth,bb=10 0 540 350,clip]{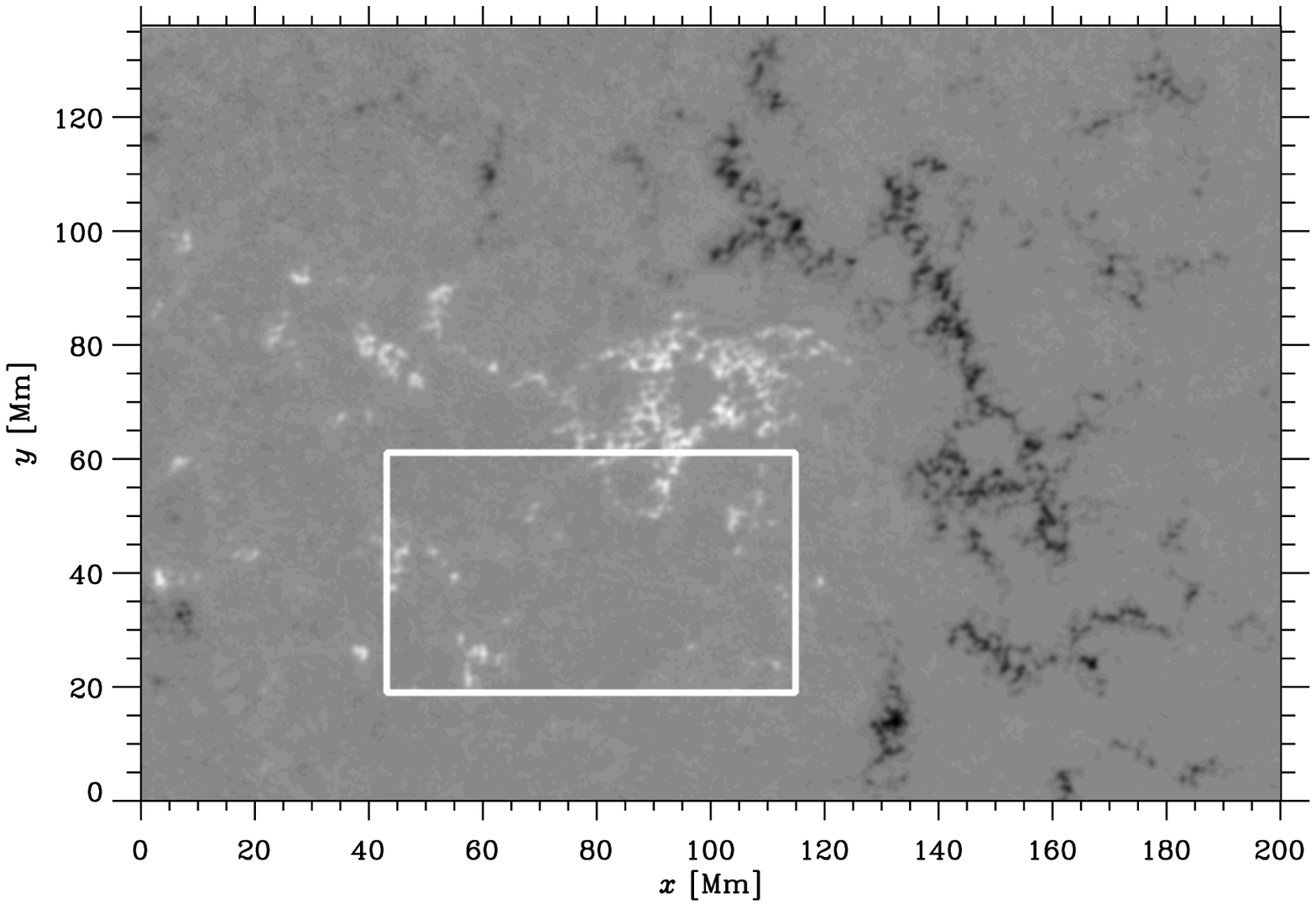}
\includegraphics[width=0.35\textwidth,bb=10 0 540 350,clip]{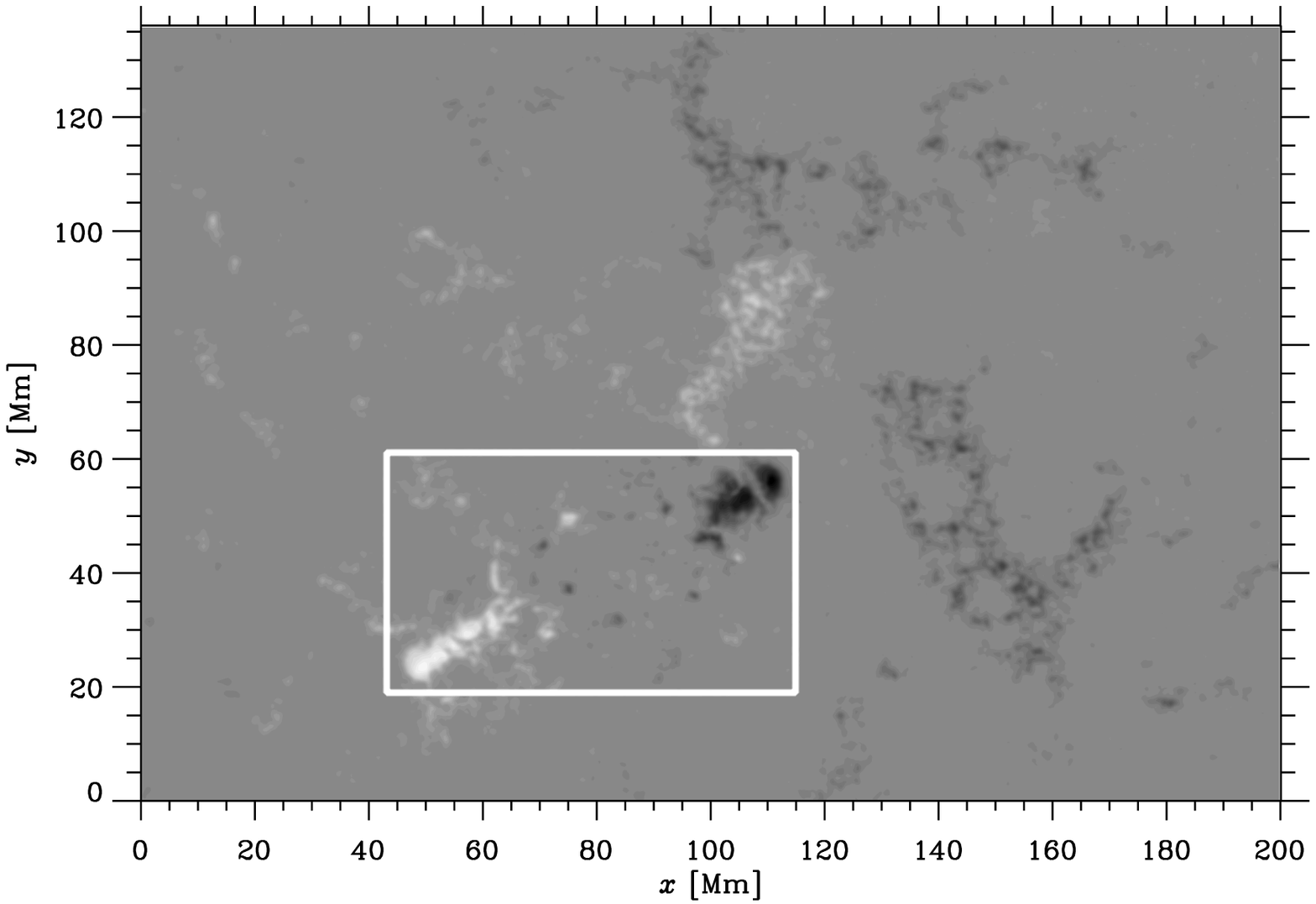}
\caption{Maps of the vertical magnetic-field component for 2016 May 22, 23:00 (left) and 2016 May 25, 08:00 (right). The white frame delineates the area for which the extrema are taken and the fluxes are calculated.}
\label{area}
\end{figure*}

\begin{figure*}
\centering
\includegraphics[width=0.35\textwidth,bb=12 10 550 350,clip]{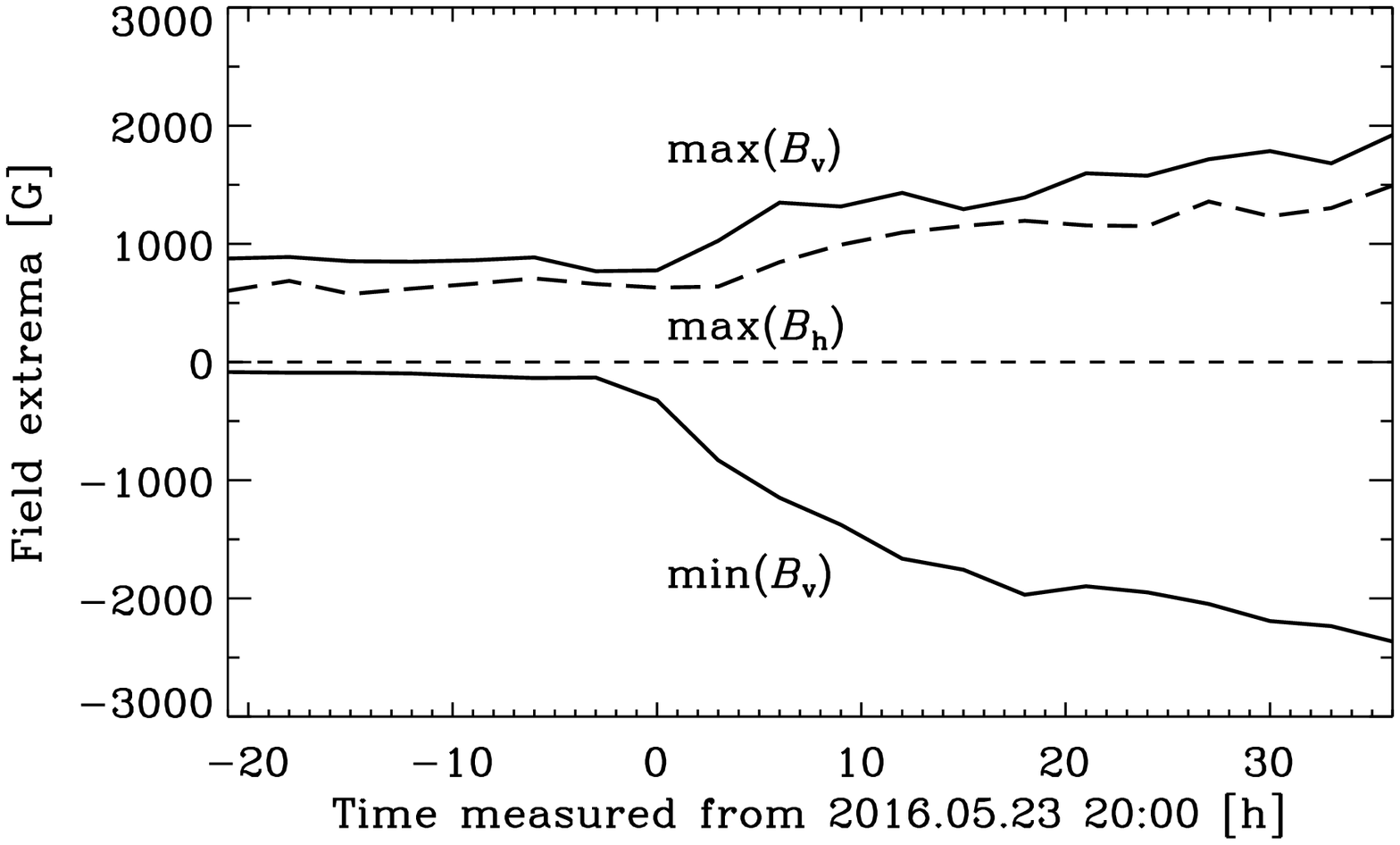}
\includegraphics[width=0.35\textwidth,bb=12 10 550 350,clip]{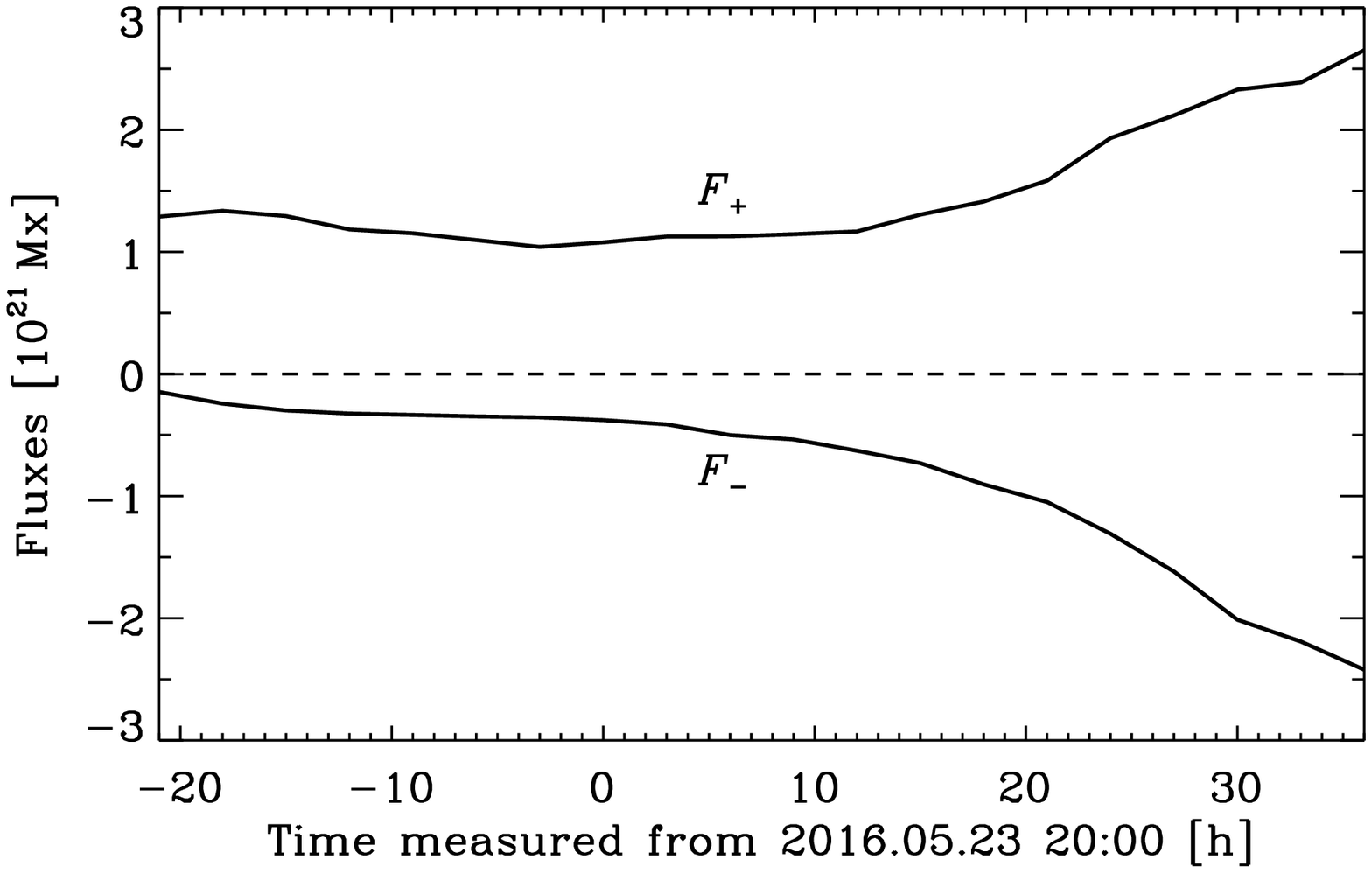}
\caption{Time variation of the magnetic-field extrema (left) and the positive and negative fluxes (right) for the area marked in Figure~\ref{area}.}
\label{ExtrFlux}
\end{figure*}

\subsection{Evolution of the Magnetic Field}

The evolution of the magnetic field in the growing BMR at the early development stage of the AR under study is illustrated in the left column of Figure~\ref{images_vs_profiles}. A map of the vertical magnetic field, $B_\mathrm v$, for the RT and four full-vector magnetic-field maps for four subsequent times are shown. We do not show the horizontal component of the field in the first map (for the RT) to clearly indicate a line segment assumed to be the BMR axis. \textbf{For each time, we draw such an axis approximately through the centroids of the main magnetic elements of the BMR. The very small differences between the line segments thus obtained are due to deformations of the magnetic-element areas.}  In the right column of the same figure, profiles of $B_\mathrm v$ variation along this axis are given for the respective times \textbf{together with similar profiles of the longitudinal field, $B_\mathrm l$ -- the projection of the magnetic field onto the BMR axis, and the vertical velocity-field component, $u$}.

It can be seen that, at the RT, weak diffuse magnetic fields with predominantly positive vertical component (i.e., of the trailing polarity) and a few small magnetic elements, in some cases corresponding to pores, occur over the whole SHARP. At the future location of the BMR, there are two magnetic elements of the positive (trailing) polarity, which are not yet associated with pores. They can clearly be identified in the $B_\mathrm v$ profile for the RT as two peaks with amplitudes of about 600~G.

At 20:12 TAI, these two positive magnetic elements are still present (and, at their locations, two pores are now distinguishable; they can be seen in the plate of Figure~\ref{white} for 23:00). However, in an enlarged $B_\mathrm v$ map (not presented here), an extremely faint shadow of the leading (negative) polarity closely adjacent to the positive element that occupies a leading position can be noticed for the first time. As the profile for 20:12~TAI in Figure~\ref{images_vs_profiles} shows, this shadow can be associated with a very shallow minimum of $B_\mathrm v$ located in the immediate neighbourhood of the positive-polarity magnetic element occupying the leading position (the local $B_\mathrm v$ extrema in the magnetic elements may not be located exactly on the line segment chosen as the BMR axis, which is why the minimum of $B_\mathrm v$ is almost imperceptible in the profile for time 20:12~TAI). This minimum becomes deeper and forms a distinct leading-polarity magnetic element by 20:48~TAI, after which the neighbouring local maximum (i.e., the positive-polarity element that was originally present and had the leading position) disappears within an hour. The growing negative (leading) element of the BMR remains in close contact with the ``old'' positive element as long as the latter exists. By 22:48~TAI, both the leading negative and trailing positive $B_\mathrm v$ extrema become comparable in magnitude, a well-defined BMR has formed, and its magnetic elements are related to a bipolar couple of pores (see Figure~\ref{white}), which subsequently develops into a bipolar sunspot group.

A consideration of the maps in the left column of Figure~\ref{images_vs_profiles} indicates that, quite expectedly, the vectors of the horizontal magnetic-field component, $\mathbf{B}_\mathrm h$, diverge from the trailing-polarity elements (where $B_\mathrm v>0$) and converge to the leading-polarity elements (where $B_\mathrm v<0$). In the growing BMR, this
convergence becomes progressively more pronounced with the formation of the leading-polarity element. The magnetic field directed from the trailing to the leading element is smeared over some area, and its characteristic values can be inferred from the longitudinal field, $B_\mathrm l$.

The behavior of $B_\mathrm l$ deserves a special discussion, since it is directly related to the expectable implication of the rising-tube process noted as feature 2 in the Introduction. To this end, we present the profiles of $B_\mathrm l$ variation along the BMR axis \textbf{in the right column of Figure~\ref{images_vs_profiles}, on the same panels where the profiles of $B_\mathrm v$ are shown}. It is remarkable that $|B_\mathrm l|$ in between the two magnetic poles of the BMR is typically below a level of 200~G. This field achieves considerably larger magnitudes only after the formation of the BMR in the neighbourhood of its magnetic poles (passing through zero exactly at the poles). Therefore, it demonstrates the feature noted in Paper II as the bordering effect: it reaches two extrema, opposite in sign, on both sides of either extremum of $B_\mathrm v$; in maps of $B_\mathrm h$, which we do not present here for the AR at hand but have presented in Paper II for another AR (in Figures~2--5, left), this feature appears as a segment of a bright ring bordering the dark central spot, where $|B_\mathrm v|$ is large and $B_\mathrm h$ is small. This reflects the fountainlike spatial configuration of magnetic field lines, which are mainly vertical in the center of the magnetic element and diverge around the center above, progressively inclining with the distance from the center. There are no signs of strong horizontal magnetic field between the future pole positions, which would be indicative of the emergence of the flux-tube-loop apex.

\begin{figure*}
\centering
\includegraphics[width=0.43\textwidth,bb=0 0 435 360,clip] {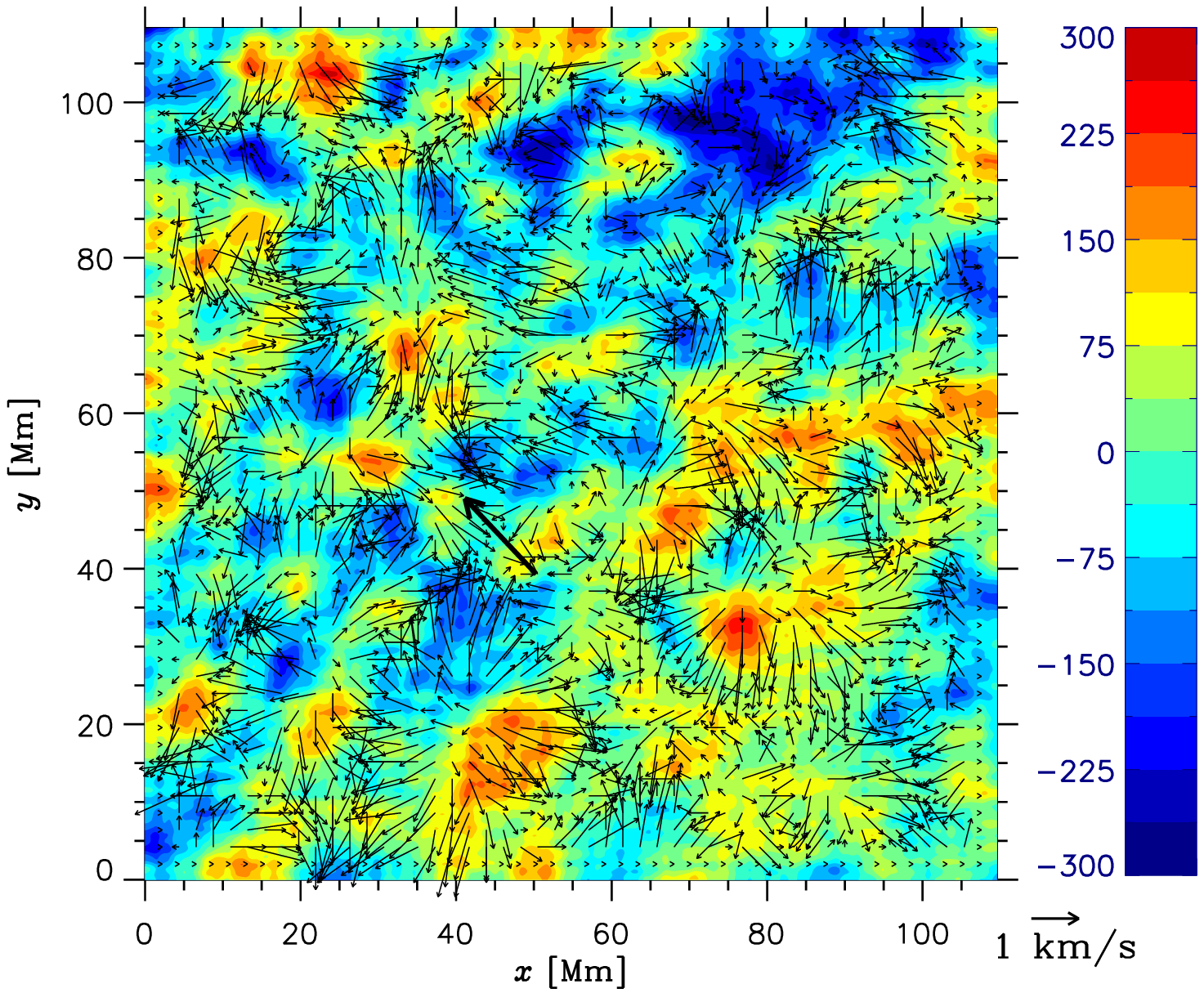}
\includegraphics[width=0.43\textwidth,bb=0 0 435 360,clip] {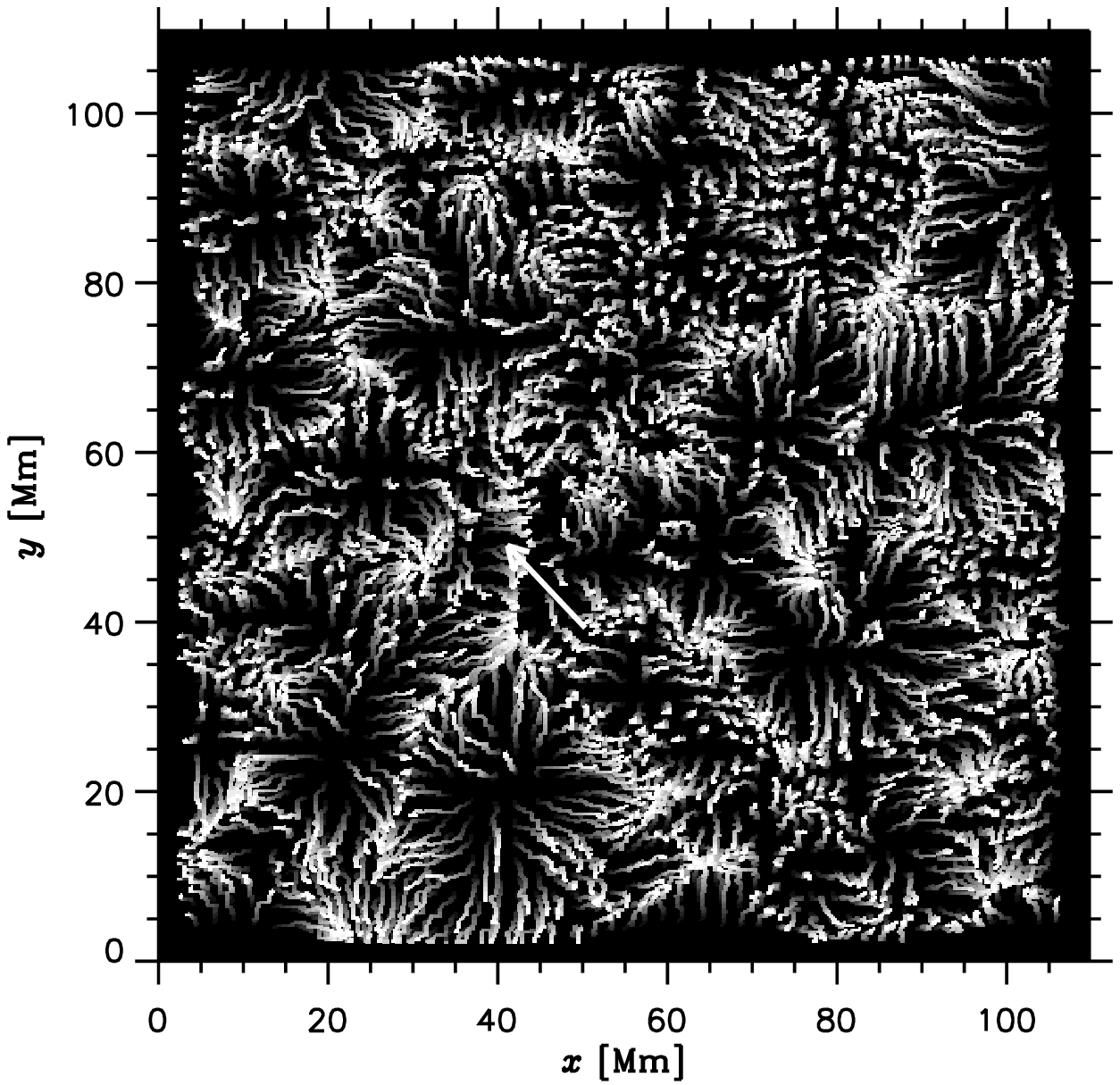}\\
\tiny{\hspace{0.7cm}2016 May 23, 20:25:30 TAI}\\[6pt]
\includegraphics[width=0.43\textwidth,bb=0 0 435 360,clip] {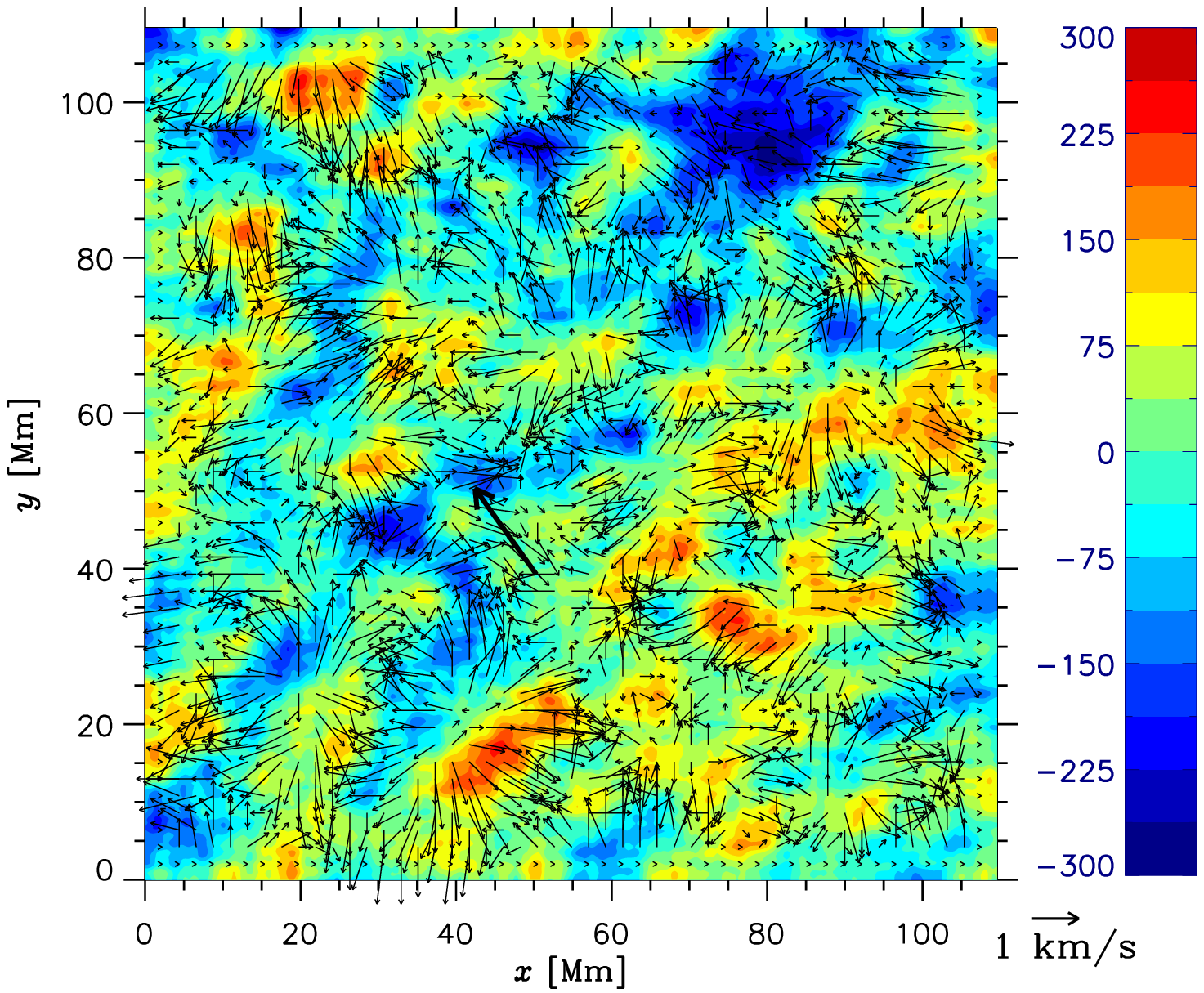}
\includegraphics[width=0.43\textwidth,bb=0 0 435 360,clip] {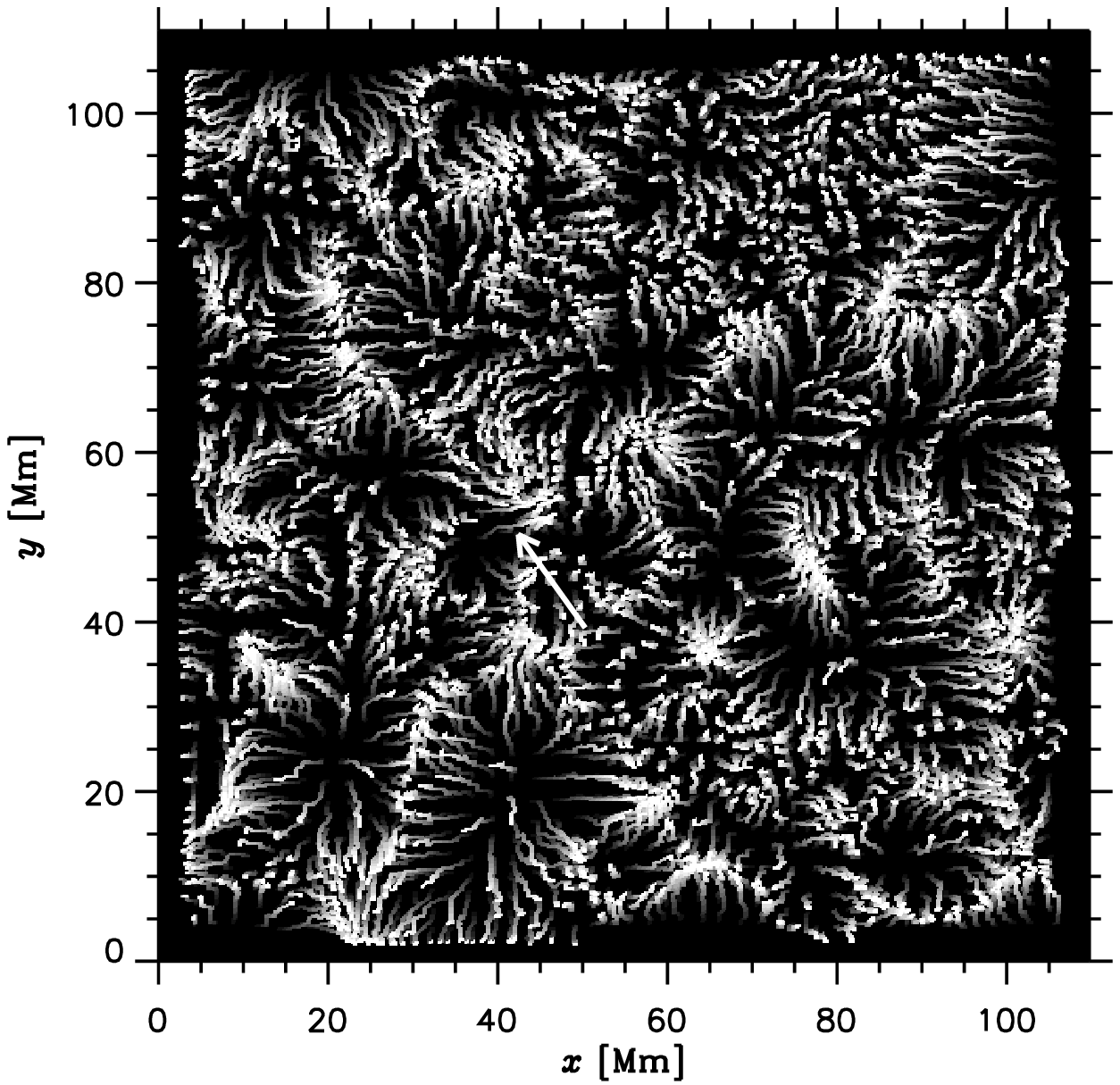}\\
\tiny{\hspace{0.7cm}2016 May 23, 21:37:30 TAI}\\[6pt]
\includegraphics[width=0.43\textwidth,bb=0 0 435 360,clip] {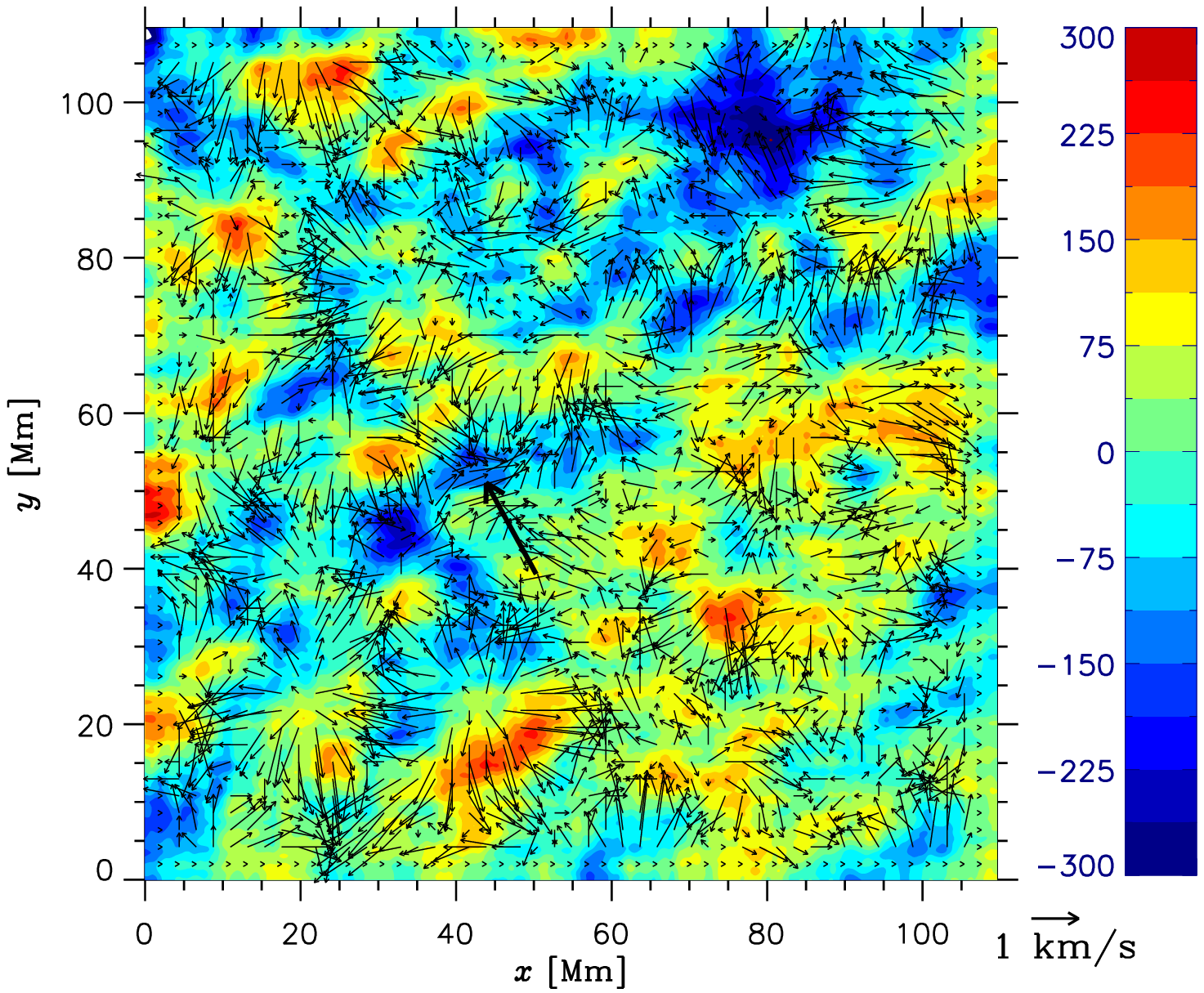}
\includegraphics[width=0.43\textwidth,bb=0 0 435 360,clip] {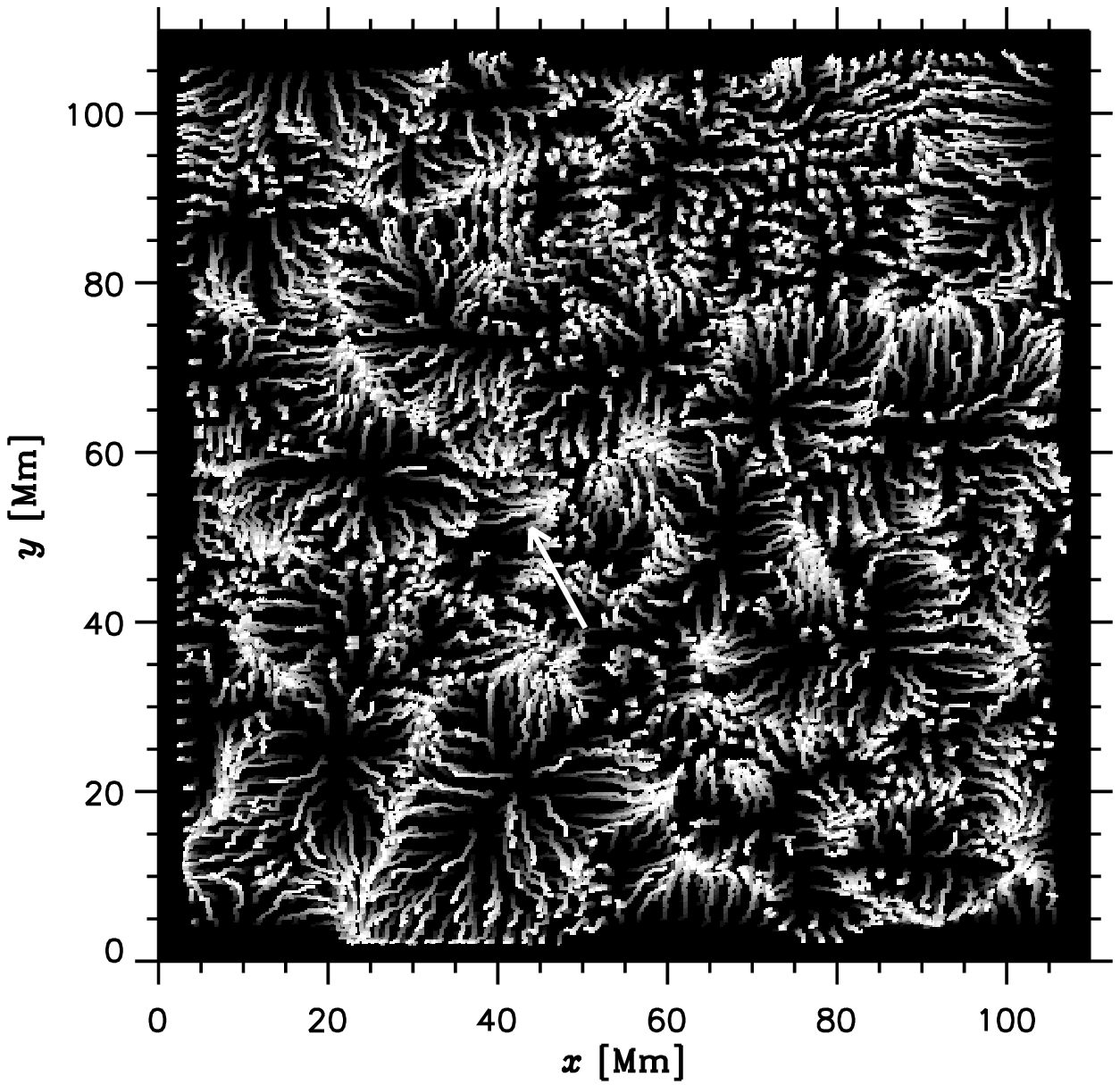}\\
\tiny{\hspace{0.3cm}2016 May 23, 22:24:45 TAI}\\[6pt]
\caption{Velocity field in the area where the BMR originates (note that both the frame position and the coordinate origin differ from those in the other figures). Left: \mbox{full}-velocity maps, the colours representing the vertical velocity component (in $\mathrm {m\,s}^{-1}$) for three selected times (indicated under each row of maps) and the arrows representing the horizontal velocity component (averaged over an interval of 20~min 15~s started from that times); right: maps of the corks trajectories obtained by integrating their displacements over an interval of 2~h~17~min centered at the respective selected times (the initial point of each trajectory is black and the final point is white, the brightness gradually increasing with time). The heavy arrows (black in the left and white in the right column) indicate the position of the leading-polarity nucleus at the corresponding times.}
\label{vel}
\end{figure*}

The time variations of the amplitude magnetic-field values and magnetic flux are descriptive of the BMR-evolution process. To obtain characteristics of the sort, we selected an area, in which dramatic changes in the magnetic-field pattern occur, and chose a time interval somewhat longer than that considered up to now, starting 21~h before and ending 36~h after the RT (Figure~\ref{area}). As can be seen from the left panel of Figure~\ref{ExtrFlux}, the positive and negative extrema of $B_\mathrm v$ vary in considerably different ways. The growth of the leading-polarity (negative) field described as the variation in $\min (B_\mathrm v)$ sets in quite abruptly near the RT (zero time in the graph) from values of about 100~G, indicating that the formation of the BMR begins, and is more rapid. The magnitude of this quantity changes by a factor of about 25 in less than 40~h. In contrast, $\max (B_\mathrm v)$ is initially of order 1000~G, varies more smoothly, and this trailing-polarity (positive) field proves to be amplified only by a factor of about 2.

In addition to the strong dissymmetry between the leading-polarity and trailing-polarity evolution, we have to note a remarkable feature of the variations in $\max B_\mathrm h$ (the dashed curve in the left panel of Figure~\ref{ExtrFlux}). In a similar way to $\max B_\mathrm v$, this quantity does not exhibit dramatic changes in the rate of its variation, also growing by a factor of about 2. Thus, no signs of the emergence of a flux tube -- the process of which the variation curve should be indicative -- can be noted.

The dissymmetry between the leading and trailing polarities in their evolution can also be clearly seen in the variation of the magnetic fluxes of either sign (the right panel of Figure~\ref{ExtrFlux}). On the whole, the pattern of variation of the magnetic flux is similar to that of the $B_\mathrm v$ extrema, although the onset of the BMR formation is not so pronounced in the flux variation. In the time interval considered, the total positive magnetic flux through the selected area, $F_\mathrm +$, changes by a factor of about 25, while the negative flux, $F_\mathrm -$, has approximately doubled.

In terms of the behavior of the magnetic polarities, this AR is in striking contrast to, e.g., the ARs described by \cite{Centeno2012}, where the positive and negative fluxes are very well balanced during the first 15~h of the BMR development.

\vspace{12pt}
\subsection{The Behavior of the Velocity Field}\label{velfield}

In view of evaluating the applicability of the RTM to the origin of the AR under study, it is instructive to consider the profiles of variation of the vertical velocity, $u$, along the BMR axis at different times (see again the right column of Figure~\ref{images_vs_profiles}), and this is worth doing in comparison with the profiles of $B_\mathrm v$ variation (the right column of Figure~\ref{images_vs_profiles}). Remember that the $B_\mathrm v$ profile exhibits two well-defined extrema (magnetic elements) starting, roughly speaking, from time 21:48~TAI; they are located at $l\approx 2.5$ and 15 ($l$ being the coordinate measured along the BMR axis). At the RT (20:48~TAI), the $u$ profile indicates the presence of two pronounced vertical flows, an upflow and a downflow. Both of them are in between the future positions of the magnetic elements but the downflow almost coincides with the location where the leading polarity will appear. By 21:48~TAI, the upflow has degenerated into a fairly narrow and weak stream, making room for a wider downflow at $l\lesssim 9$,  while the downflow at the location of the leading-polarity element ($l\approx 15$) still exists. At later times (e.g., 22:48~TAI), there are three downflows and two upflows at the BMR axis. It can therefore be concluded that no predominant upflow precedes the origin of the BMR. Generally, upflows and downflows are mixed, with some prevalence of downflows.

Now let us consider the entire pattern of the full-vector velocity field in the surroundings of the growing BMR. Figure~\ref{vel} shows this field for three selected times: shortly after the RT and about one and two hours later. It can easily be seen from the left column of panels that both the vertical and the horizontal velocity field are distributed very similarly at all these times. In particular, these maps confirm our inference that there is no upflow dominating in the area where the BMR forms; moreover, downflows even prevail in this area. Another important feature is the absence of any signs of spreading, or HDF, from the emergence area of the BMR. In contrast, as demonstrated by the maps of cork trajectories traced over an interval of 2~h 17~m (right column), the pattern of regular supergranules and mesogranules is preserved in the horizontal-velocity field; it varies little during the two-hour interval (this pattern \textbf{being virtually the same but even more} pronounced if the cork displacements are integrated over a 4-h interval; \textbf{we do not present this map here}). The accumulation of corks at the cell boundaries outlines the supergranulation and mesogranulation network and emphasizes its stability.

In the context of the observed horizontal velocities, it is worth mentioning again the study by \citet{Birch_etal_2016}. They tried to reveal HDFs deriving horizontal velocities from SDO/MHI observations of the solar surface around emerging active regions and using in parallel their numerical simulations of solar magnetoconvection in the presence of an emerging model flux tube. For 70 ARs considered, the one-$\sigma$ range of azimuthally averaged radial-outflow speeds at a distance of 15~Mm from the expected emergence location, at 3 hours before the emergence time, was found to be $-8 \pm 50\ \mathrm{m\,s}^{-1}$, while the similar range for quiet-Sun regions chosen for control purposes was $-5 \pm 40\ \mathrm{m\,s}^{-1}$. If the rising-tube mechanism is assumed, the observed flow patterns can be associated with tube-rise speeds not exceeding 150 m\,s$^{-1}$ at a depth of 20 Mm. This figure agrees with the estimated convection velocities at this depth but is well below the prediction of the emerging-flux-tube model. The authors conclude that the dynamics of the emerging magnetic field in the subphotospheric layers is controlled by convective flows.

\section{Summary of Results}

The following remarkable traits are characteristic of the origin and early development stage of AR~12548 considered here:
\begin{enumerate}
\item The leading-polarity (negative) magnetic element of the BMR originates as a compact feature with a fountainlike magnetic-field structure against the background of a distributed trailing-polarity field, in which a nucleus of the trailing-polarity (positive) element is already present. The negative element is in close contact with another pre-existing positive element, which subsequently disappears.
\item There are no signs of a strong horizontal magnetic field between the nuclei of the magnetic poles of the BMR, which would indicate the emergence of the apex of an intense magnetic-flux tube. The horizontal magnetic field does not exhibit dramatic changes. Immediately before the origin of the BMR and during its early development stage, the projection of the magnetic-field vector onto the BMR axis is typically below 200~G in between the future positions of the two magnetic poles, thus being not associated with the emergence of a strong flux tube.
\item No predominant upflow between the future locations of the magnetic poles precedes the origin of the BMR. Instead, upflows and downflows are mixed, and downflows even prevail in this area. The leading-polarity magnetic element nucleates against the background of a downflow.
\item There are no signs of large-scale spreading, or HDF, from the area where the BMR develops. In contrast, a regular supergranulation and mesogranulation pattern remains intact.
\item There is a strong dissymmetry between the time variations of the negative and positive extrema of the magnetic field and between the time variations of the negative and positive magnetic fluxes through some area encompassing the BMR: the growth of the leading (negative) polarity sets in abruptly and occurs rapidly while the trailing (positive) polarity grows smoothly and more slowly. In a 57-hour interval encompassing the abrupt onset of the leading-polarity growth, the amplitude and the magnetic flux of the leading polarity increase by a factor of about 25, while those of the trailing polarity only double.
\end{enumerate}

\section{Discussion and Conclusion}

Our analysis of the data on the early development stage of AR 12548 suggests a number of conclusions concerning the phenomena involved. Items 2--4 in the above list of results -- the lack of a strong horizontal magnetic field, which should reflect the emergence of the apex of the flux-tube loop; the lack of an overall upflow on the scale of the growing AR, which should be indicative of the flux-tube emergence; and the lack of a spreading flow (HDF) around the area of the growing BMR -- are in strong contradiction with the idea of the emergence of an $\Omega$-shaped intense-flux-tube loop. It is also worth noting some other details of the process.

The pattern of the BMR development demonstrates a great dissimilarity between the leading and trailing magnetic polarities in their behavior. The leading polarity nucleates as a compact isolated feature against the background of a distributed trailing-polarity magnetic field. The growth of the leading polarity starts from noise values, which scarcely exceed 100~G -- in essence, from the complete absence of any signature of the future leading magnetic pole of the BMR. For some time, the negative (leading-polarity) magnetic element grows in close contact with a pre-existing positive element, which rapidly decays. In contrast, the starting strength of the trailing polarity is slightly below 1000~G, and the trailing magnetic pole develops as a ``condensation'' of the pre-existing background field. Both of these radically different scenarios appear to be hardly compatible with the notion of the emergence of an $\Omega$-shaped loop.

The persistence of the supergranulation and mesogranulation pattern during the formation of the BMR brings back memories to the observations reported many years ago by \citet{bum63,bum} and \citet{bumhow}. According to these researchers, the growing magnetic fields of BMRs do not break down the pre-existing convective-velocity field but come from below ``seeping''   through the network of convection cells.

Thus, our principal conclusion is the inconsistency of the scenario of the origin and early evolutionary stage of AR~12548 with the idea of emergence of an $\Omega$-shaped flux-tube loop carrying a strong magnetic field. We were able to catch the origin of the BMR within several minutes and keep track of the process in its most refined (``naked'') appearance, without interference from other magnetic features complicating the pattern. The observed scenario suggests that an \emph{in situ} mechanism should operate in this case, and plasma motion rather than the magnetic field seems to be basically responsible for the formation of the BMR.

The BMR-development pattern in AR~12548 should not necessarily be typical of most ARs. Nevertheless, both our case studies -- that described in Papers I and II and especially the present one -- clearly indicate that the RTM offers by far not a universal possibility of AR and sunspot-group formation.

Gathering observational data and systematizing various evolutionary scenarios of AR formation appear to be necessary to comprehend the complex of physical mechanisms responsible for the development of solar-activity processes in the convection zone and atmosphere of the Sun. The above-described study can be considered a particular contribution to the implementation of this general program. The currently available abundant and detailed observational data for AR dynamics offer possibilities for an enormous extension of the scope of studies similar to the present one. We plan further steps on this avenue, and the elaborated techniques of data processing are a favourable prerequisite for such investigations.

\begin{acknowledgements}
The observational data were used here by courtesy of NASA/SDO and the HMI science teams. The kind assistance by Arthur Amezcua, Philip Scherrer, Todd Hoeksema and Xudong Sun in dealing with HMI data available via JSOC is gratefully acknowledged.
\end{acknowledgements}

\bibliography{Getling}

\end{document}